\documentclass[aps,prb,superscriptaddress,shortbibliography,singlecolumn]{revtex4-2}

\usepackage{lmodern}

\usepackage[latin9]{inputenc}
\usepackage{amsmath}
\usepackage{amssymb}
\usepackage{graphicx}
\usepackage{epstopdf}
\usepackage{color}
\usepackage{enumerate}
\usepackage{ulem}
\usepackage{caption}

\usepackage{chemformula}

\begin{document}

\title{Linear colossal magnetoresistance driven by magnetic textures in \ch{LaTiO_3} thin films on \ch{SrTiO_3}}

\date{\today} 					

\author{Teresa Tschirner}
\affiliation{Leibniz Institute for Solid State and Materials Research, IFW Dresden, Helmholtzstr. 20, 01069 Dresden, Germany}
\affiliation{W\"urzburg-Dresden Cluster of Excellence ct.qmat, Germany}
\author{Berengar Leikert}
\affiliation{Physikalisches Institut, Universit\"at W\"urzburg, D-97074 W\"urzburg, Germany}
\affiliation{W\"urzburg-Dresden Cluster of Excellence ct.qmat, Germany}
\author{Felix Kern}
\affiliation{Leibniz Institute for Solid State and Materials Research, IFW Dresden, Helmholtzstr. 20, 01069 Dresden, Germany}
\author{Daniel Wolf}
\affiliation{Leibniz Institute for Solid State and Materials Research, IFW Dresden, Helmholtzstr. 20, 01069 Dresden, Germany}
\author{Axel Lubk}
\affiliation{Leibniz Institute for Solid State and Materials Research, IFW Dresden, Helmholtzstr. 20, 01069 Dresden, Germany}
\affiliation{Institute of Solid State and Materials Physics, TU Dresden, 01069  Dresden,  Germany}
\affiliation{W\"urzburg-Dresden Cluster of Excellence ct.qmat, Germany}
\author{Martin Kamp}
\author{Kirill Miller}
\author{Fabian Hartmann}
\author{Sven H\"ofling}
\affiliation{Physikalisches Institut, Universit\"at W\"urzburg, D-97074 W\"urzburg, Germany}
\affiliation{W\"urzburg-Dresden Cluster of Excellence ct.qmat, Germany}
\author{Bernd B\"uchner}
\affiliation{Leibniz Institute for Solid State and Materials Research, IFW Dresden, Helmholtzstr. 20, 01069 Dresden, Germany}
\affiliation{Institute of Solid State and Materials Physics, TU Dresden, 01069  Dresden,  Germany}
\affiliation{W\"urzburg-Dresden Cluster of Excellence ct.qmat, Germany}
\author{Joseph Dufouleur}
\affiliation{Leibniz Institute for Solid State and Materials Research, IFW Dresden, Helmholtzstr. 20, 01069 Dresden, Germany}
\affiliation{W\"urzburg-Dresden Cluster of Excellence ct.qmat, Germany}
\author{Marc Gabay}
\affiliation{Laboratoire de Physique des Solides, Universit\'e Paris-Saclay, CNRS UMR 8502, 91405 Orsay, France}
\author{Michael Sing}
\author{Ralph Claessen}
\affiliation{Physikalisches Institut, Universit\"at W\"urzburg, D-97074 W\"urzburg, Germany}
\affiliation{W\"urzburg-Dresden Cluster of Excellence ct.qmat, Germany}
\author{Louis Veyrat}
\affiliation{Leibniz Institute for Solid State and Materials Research, IFW Dresden, Helmholtzstr. 20, 01069 Dresden, Germany}
\affiliation{Physikalisches Institut, Universit\"at W\"urzburg, D-97074 W\"urzburg, Germany}
\affiliation{W\"urzburg-Dresden Cluster of Excellence ct.qmat, Germany}

\begin{abstract}

Linear magnetoresistance (LMR) is of particular interest for memory, electronics, and sensing applications, especially when it does not saturate over a wide range of magnetic fields. One of its principal origins is local mobility or density inhomogeneities, often structural, which in the Parish-Littlewood theory leads to an unsaturating LMR proportional to mobility. Structural disorder, however, also tends to limit the mobility and hence the overall LMR amplitude. An alternative route to achieve large LMR is via non-structural inhomogeneities which do not affect the zero field mobility, like magnetic domains.
Here, linear positive magnetoresistance caused by magnetic texture is reported in \ch{LaTiO3}/\ch{SrTiO3} heterostructures. The LMR amplitude reaches up to 6500\% at 9T. This colossal value is understood by the unusual combination of a very high thin film mobility, up to 40 000 cm$^2$/V.s, and a very large coverage of low-mobility regions. These regions correlate with a striped magnetic structure, compatible with a spiral magnetic texture in the \ch{LaTiO3} film, revealed by low temperature Lorentz transmission electron microscopy. These results provide a novel route for the engineering of large-LMR devices.

\end{abstract}

\maketitle


\section{Introduction}

In conventional metals, the magnetoresistance (MR) commonly exhibits quadratic field dependence at low magnetic fields and saturates at high fields \cite{Abrikosov98}. This classical MR is usually limited in amplitude to a few percent at 10T \cite{Majumder}. Many effects, however, are known to impact magnetoresistance: a magnetic ground state (ferromagnetism (FM) \cite{Jin1994, Lin2016, Sinjukow2003, Jin, Subramanian, Roeder}, antiferromagnetism (AFM) \cite{Maksimovic2020, McGuire2021}), Dirac physics \cite{Wei, WangX,Zhang}, Landau levels \cite{Shoenberg, Datta}, or spatial inhomogeneities for instance \cite{Parish1, Parish2, Kozlova, Mallik2021, Song, Meng2020}, can lead to modifications of the amplitude or the magnetic field dependency of the MR. In particular, Dirac physics in topological insulators \cite{Wei, WangX,Zhang} and semimetals \cite{Yang,Xu, Husmann, Lee, Novak, Huang, Xiong, Narayanan, Feng} or spatial inhomogeneities \cite{Kozlova, Mallik2021, Song,Zhang2021} can result in linear magnetoresistance (LMR) behaviour \cite{Abrikosov98, Abrikosov, Parish1, Parish2}. Beyond its relevance for fundamental research, the non-saturating nature of LMR makes it also interesting for sensing, electronics and memory applications. In this context, systems with large LMR are highly desirable.\\
Amongst other theories, non-saturating LMR can be understood semi-classically in the Parish-Littlewood framework by a random resistor network model that mimics a disordered and strongly inhomogeneous conductor system \cite{Parish1, Parish2}. This model can be extended to systems with local low-mobility or low-carrier density regions acting as guiding centres for the carrier paths \cite{Kozlova, Mallik2021, Song}. This model predicts a LMR whose amplitude depends linearly on the carrier mobility and on the guiding center density\cite{Kozlova}.
Guiding center regions are, however, usually related to local structural defaults or disorder \cite{Kozlova,Song}, which limit the system mobility and hence the amplitude of the resulting LMR to about 200\% at 9T \cite{Kozlova, Mallik2021, Narayanan}. Other mechanisms inducing local inhomogeneities without affecting the zero-field mobility could therefore lead to a much stronger effect. In particular, magnetically textured regions could lead to local mobility inhomogeneities through magnetic scattering, while potentially retaining high mobility since the crystalline quality would not be affected. More recently, LMR has been observed together with ferromagnetic (FM) order \cite{He2021} or magnetic signatures (anomalous Hall effect \cite{Niu2021,Wang2021}), with high amplitudes of about 1000\% at 9T. However, tangible evidence of a link between magnetic structure and LMR has yet to be ascertained.
Here we report on a linear positive magnetoresistance in \ch{LaTiO3}/\ch{SrTiO3} (LTO/STO) heterostructures with high mobility (up to 40 000 cm$^2$/V.s). The MR amplitude is much higher than in most oxide systems (up to 6500\% at 9T), and falls within the range of the colossal magnetoresistance. This linear colossal magnetoresistance (LCMR) is found to be compatible with the Parish-Littlewood theory, with an extremely high coverage of low mobilities regions, between 50\% and nearly 90\%. Through low temperature transmission electron microscopy (TEM) in magnetic field, the origin of these regions is traced back to a magnetic uniaxial stripe pattern, oriented along well-defined crystallographic axes. Those stripes are compatible with spiral magnetic order in the LTO layer. Our work reveals that magnetic texture can induce linear magnetoresistance with colossal amplitude in high mobility devices such as LTO/STO heterostructures.

\section{Thin film samples}

\begin{figure}[b]
	\includegraphics[width=0.9\textwidth]{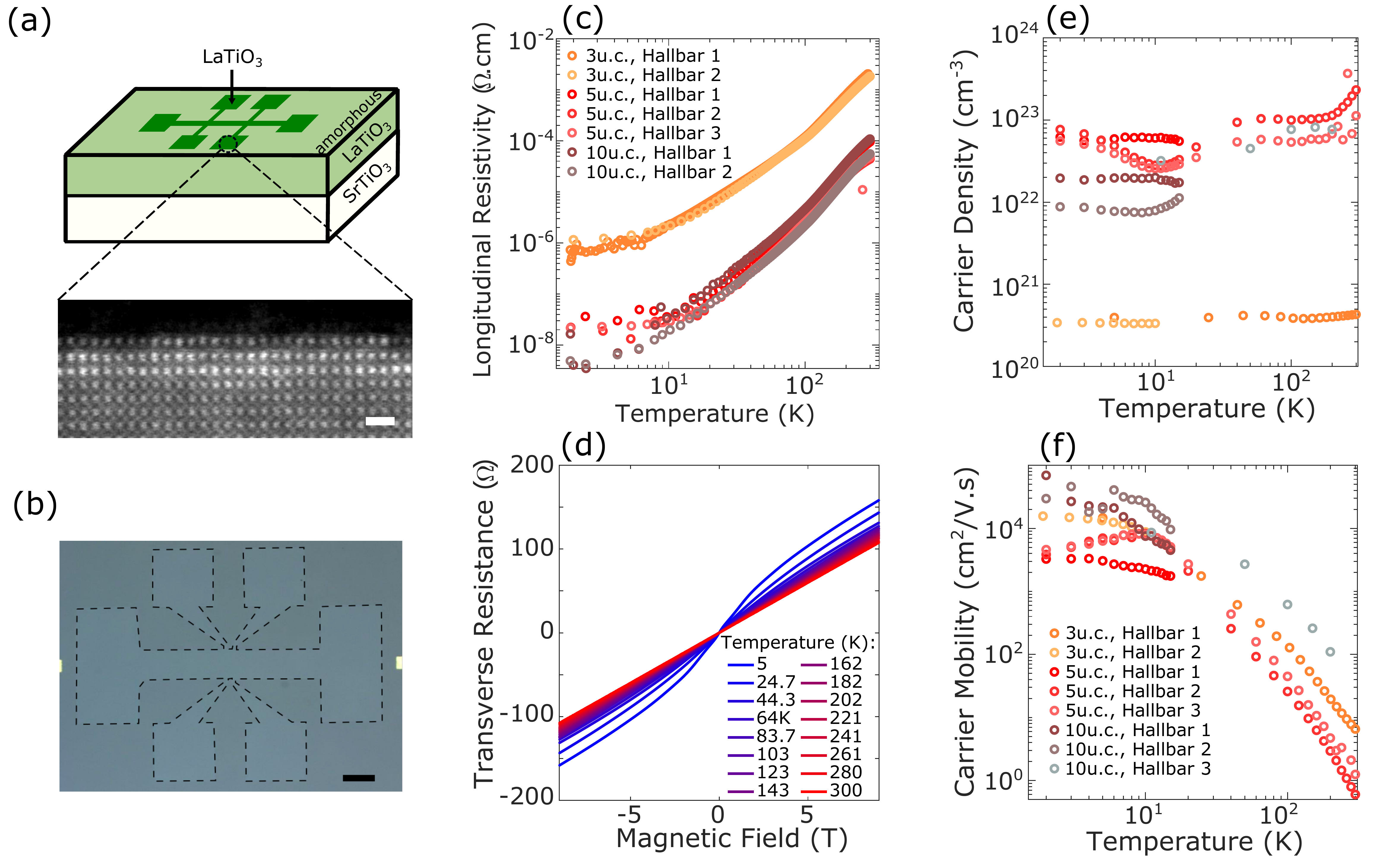}
	\caption{\textbf{(a),} Top: Schematic of the LTO thin film grown on STO in a Hall bar shape. Bottom: Cross-sectional high resolution scanning TEM image of an LTO 3u.c. (thickness 1.2~nm) thin film. Scale bar: $1~\mathrm{nm}$.  \textbf{(b),}  Optical image of a Hall bar, outlined with a dashed black line. Scale bar: 100~$\mu$m. \textbf{(c),} Resistivity versus temperature for different Hall bars on the 3, 5 and 10u.c. thick samples. \textbf{(d),} typical Hall effect measurements of the 3u.c. sample (Hall bar 1) for temperatures ranging from 5 to 300K. \textbf{(e),} Carrier density and \textbf{(f),} mobility of the 3, 5 and 10u.c. thick samples versus temperature.}
	\label{Figure1}
\end{figure}

Our LTO thin films were grown on TiO$_2$ terminated STO substrates using pulsed laser deposition. The structures where pre-patterned with optical lithography similarly to Ref.\cite{Schneider2006} to achieve crystalline LTO in Hall bar shapes as seen in Figure \ref{Figure1}(a),(b) of different lengths and widths. 

We first consider standard transport and Hall characterisation to identify the origin of the main conducting channel. Figure \ref{Figure1}(c) shows the temperature dependence of the resistivity for different film thicknesses and Hall bar geometries. The resistivity $\rho$ of different Hall bars is well reproducible for a specific film thickness with small variations accounting for spatial inhomogeneity of the films. We observe that the resistivity decreases significantly from the 3u.c. to the 5u.c. film but stays approximately constant for both the 5u.c. and the 10u.c. samples. All three investigated thicknesses exhibit clear metallic behaviour. The samples further show a high residual resistance ratio $RRR=R(T=300K)/R(T=2K)\sim1000 - 10000$, implying very clean, high quality samples with only trace amounts of impurities and structural disorder.
The Hall effect, shown in Figure \ref{Figure1}(d), is linear at room temperature, gradually evolving into an s-shape with decreasing temperature, which most probably indicates the presence of several bands contributing to transport. We extract the total carrier density $n_H$ from the high field asymptotic slope $(dR_H/dB)|_{9T}$ at 9T of the Hall resistance ($R_H$),  $n_{H}\simeq5\times10^{20} cm^{-3} $ for the 3u.c. and $\sim 1\times10^{23} cm^{-3} $ for both the 5u.c. and 10u.c. films at high temperature. These very large charge carrier densities are found to be almost temperature independent as shown in Figure \ref{Figure1}(e). The low temperature variations arise as the Hall effect starts to display non-linearities. Interestingly, the transport mobilities (Figure \ref{Figure1}(f)) extracted from the Drude formula as $\mu=\sigma (dR_H/dB)|_{9T}$ (where $\sigma$ is the conductivity) are found to be up to 40 000 cm$^2$/V.s at low temperatures reflecting the high sample quality of our thin films.

Next we discuss the spatial location of the conduction electrons in our LTO/STO heterostructures for which three distinct conduction channels are possible: through an interface 2D electron gas (2DEG)  between LTO and STO, through unintentionally doped STO, and through bulk LTO. First, the extremely high equivalent 2D carrier density of $\sim 10^{15}$cm$^{-2}$ extracted from the Hall measurements for the 5u.c. and 10u.c. samples makes a pure 2DEG scenario highly unlikely. Typical carrier densities in STO-based 2DEGs are usually one to two magnitudes smaller \cite{Biscaras, Niu2021}.  This conclusion is corroborated by the absence of a superconducting transition down to 0.2 K (data not shown), which has previously been reported for doped STO as well as for an interface 2DEG in LTO/STO \cite{Lebedev2020, Biscaras}. On the other hand, characterization of the Titanium valence in the LTO film by x-ray photoelectron spectroscopy (XPS) reveals a strong 4+ signal with a sizable 3+ admixture (see Supp. Info. I). This places the doping state of the LTO film far from the nominal Mott insulator (expected for a purely Titanium 3+ valence) and well into the metallic regime \cite{Scheiderer}. We therefore believe that the transport is dominated by the metallic bulk of the LTO films.

\section{Linear Colossal Magnetoresistance}
\label{sec:others}

\begin{figure}
	\includegraphics[width=1\textwidth]{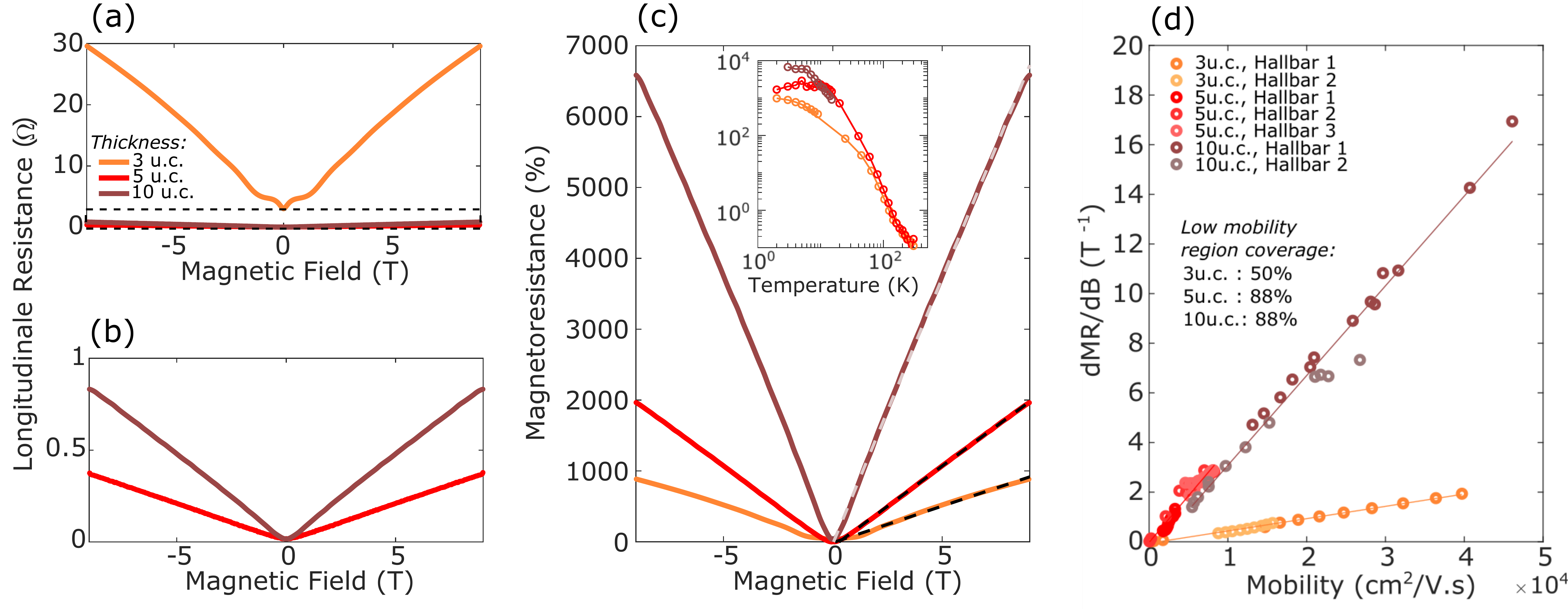}
		\caption{\textbf{Magnetoresistance as a function of the perpendicular magnetic field for LTO thicknesses of 3, 5, and 10u.c.} \textbf{(a)}, Linear magnetoresistance at T=3K for all three thicknesses and \textbf{(b)}, a zoomed in view of only the 5u.c. and the 10u.c. samples. \textbf{(c)}, Magnetoresistance MR (as defined in the main text) of the samples shown in \textbf{(a)} plotted in \%. The dashed lines corresponds to a linear fit. Inset: the temperature dependence of the MR in a log-log scale for the three thicknesses at B=9T. \textbf{(d)}, linear slope of the MR at 8T as a function of the mobility of LTO thin films of different thicknesses. Solid lines correspond to linear fits. The areal coverage (in \%) of the low mobility regions are extracted from the linear fit.}
	\label{Figure2}
\end{figure}

We now turn to an analysis of the magnetoresistance. Figure \ref{Figure2}(a) shows the dependence of the longitudinal resistance as a function of the perpendicular magnetic field for Hall bars of different thicknesses studied at 3K. The 3u.c. sample displays a strong positive MR, and further exhibits low-field features that point to Sondheimer oscillations (see Supp. Info. II). The relative increase is even stronger for the 5 and 10u.c. samples as shown in Figure \ref{Figure2}(b). Furthermore, it can be seen from these measurements that the response is not just colossal but also linear with remarkable precision, starting at magnetic fields well below 1T and with no visible saturation up to at least 11T. All samples studied systematically exhibit this non-saturating linear magnetoresistance.
To better display the relative increase of the resistance, we plot the MR in \% calculated as:
$$MR(\%)=\frac{R(B)-R(B=0)}{R(B=0)} \times 100$$
     
MR amplitudes of more than 700\% at B=9T for the 3u.c. and up to 6500\% at B=9T for the 10u.c. film are observed, which fall within the range of the colossal magnetoresistance \cite{Ramirez}. The MR therefore increases sizably with increasing thickness going from the 3u.c. to 10u.c film. In addition the linear regime sets in for field values that get smaller as the thickness is increased (see Supp. Info. III). This thickness evolution corroborates our conclusion that the transport involves carrier motion in the LTO film.

In the inset of Figure \ref{Figure2} we plot the MR at 9T versus temperature in a log-log scale. The MR amplitude shows a saturation below 10K and a strong decrease with increasing temperature. In addition, a departure from the linearity is measured for the magnetoresistance and becomes more parabolic-like above $\sim60-80$K (see Supp. Info. III).

We see a marked contrast between our findings and those reported in Ref.\cite{Veit} which showed the appearance of Shubnikov-de-Haas oscillations starting around 1T and non-linear magnetoresistance of at most 400\% at 9T \cite{Veit}. The origin of this discrepancy is unclear. Our observation of an LMR is, however, fully reproducible over several samples of varying thickness and several Hall bar geometries. The absence of Shubnikov-de-Haas oscillations in our samples despite large transport mobilities, which should allow their observation in our field range, could point to inelastic scattering in our films due to strong spin-orbit coupling. This effect can cause the quantum mobility (which determines the onset of the Shubnikov-de-Haas regime) to be reduced compared to the transport mobility \cite{Dufouleur2016}.

The occurrence of LMR has been previously observed in studies on other materials, albeit with significantly reduced amplitude as compared to our results \cite{Lonchakov2021, Jin2016, Kormondy2018}. The scenarios that have been proposed to explain the LMR involve either quantum or classical effects. In the case of a quantum origin \cite{Abrikosov98}, the linear MR appears in the limit, when all the electrons occupy the lowest Landau level. The MR is calculated as $\rho_{xx}=N_i B/\pi n^2e$, with $n$ being the carrier density and $N_i$ the concentration of scattering centers. This expression is valid for $B<(\hbar/e)n^{3/2}$ and with an electron density of $\sim2\times10^{23} cm^{-3} $, the field $B_z=\hbar (3\pi^2 n_e)^{2/3}/2e$ at which electrons coalesce into a single Landau level is $1 \times 10^5$ T  \cite{Kozlova}. This is far greater than the values in our systems which show an onset well below 1T.

The classical model, on the other hand, describes the origin of the LMR as distortions in the current paths that can be induced by microscopic spatial fluctuations in the carrier mobility \cite{Parish1}. Such a classical linear MR has been observed in highly disordered systems \cite{Hu} as well as in high mobility systems with weak disorder \cite{J.Wang} where the carrier concentration is too high to freeze the electrons in the lowest Landau level (quantum limit). In this model the system is made of low-mobility regions embedded in a high mobility conductor. The stochastic behaviour of the electron trajectories around the low-mobility regions induces the LMR \cite{Kozlova, Patane, Song}. In this theory, at high magnetic field the amplitude of the LMR varies linearly with the average global mobility (determined at zero field), as given by the following equation:
\begin{equation}
MR=\frac{s}{2L} \mu B
\label{eqn1}
\end{equation}
where $s$ is the average radius of the regions, $L$ is the distance between the regions and $\mu$ is the mobility.

To validate this scenario in our system, we have plotted the linear slope of the MR curve with respect to the mobility. Thanks to the strong temperature dependence of the Hall mobility in our samples, we can observe the dependence of the LMR slope over a broad mobility range (Figure \ref{Figure2}(d)).
It is clearly seen that the LMR slope depends linearly on the Hall mobility as expected from Eq. (\ref{eqn1}). Moreover, different Hall bar geometries with identical thickness consistently fall on the same curve, indicating perfect reproducibility between different samples, also with respect to their inhomogeneity. This linear dependence confirms the presence of low-mobility regions as the origin of the LMR with the mobility in our samples. Using equation (\ref{eqn1}), we can extract the coverage $s/(s+L) = \frac{1}{1+L/s}$ of these low-mobility islands via the dependence of the slope of the LMR on the mobility ($\mu$). We obtain coverages ranging from 50\% for the 3u.c. sample to 88\% for both the 5 and the 10u.c. samples. This is 2 to 6 times larger than coverages reported so far \cite{Mallik2021, Kozlova}.

The amplitudes of the LMR in our samples are very large for systems with such high coverages of low-mobility regions. The reason for this lies in the combination of both the high background mobility and the substantial coverage of low-mobility regions. As can be seen from Eq.~(\ref{eqn1}), in the guiding centres model, the amplitude of the LMR is proportional to both Hall mobility ($\mu$) and region coverage. In standard thin films and 2DEGs, low-mobility regions usually originate from local disorder, whether structural or induced by impurities \cite{Kozlova}. As such, an increase in the Hall mobility involves a decrease in the island concentration and conversely. This results in a balance which tends to limit the LMR amplitude. In our system however, we observe both a very high mobility and an exceptional coverage of low-mobility regions. This unusual combination allows us to reach LMR with colossal amplitudes.

However, the combination of very high mobility and large coverage of guiding centres seems incompatible with a structural origin of the low-mobility islands. In the next section, the nature of those low-mobility regions is discussed. 

\section{Magnetic stripe pattern}

\begin{figure}
	\includegraphics[width=1\textwidth]{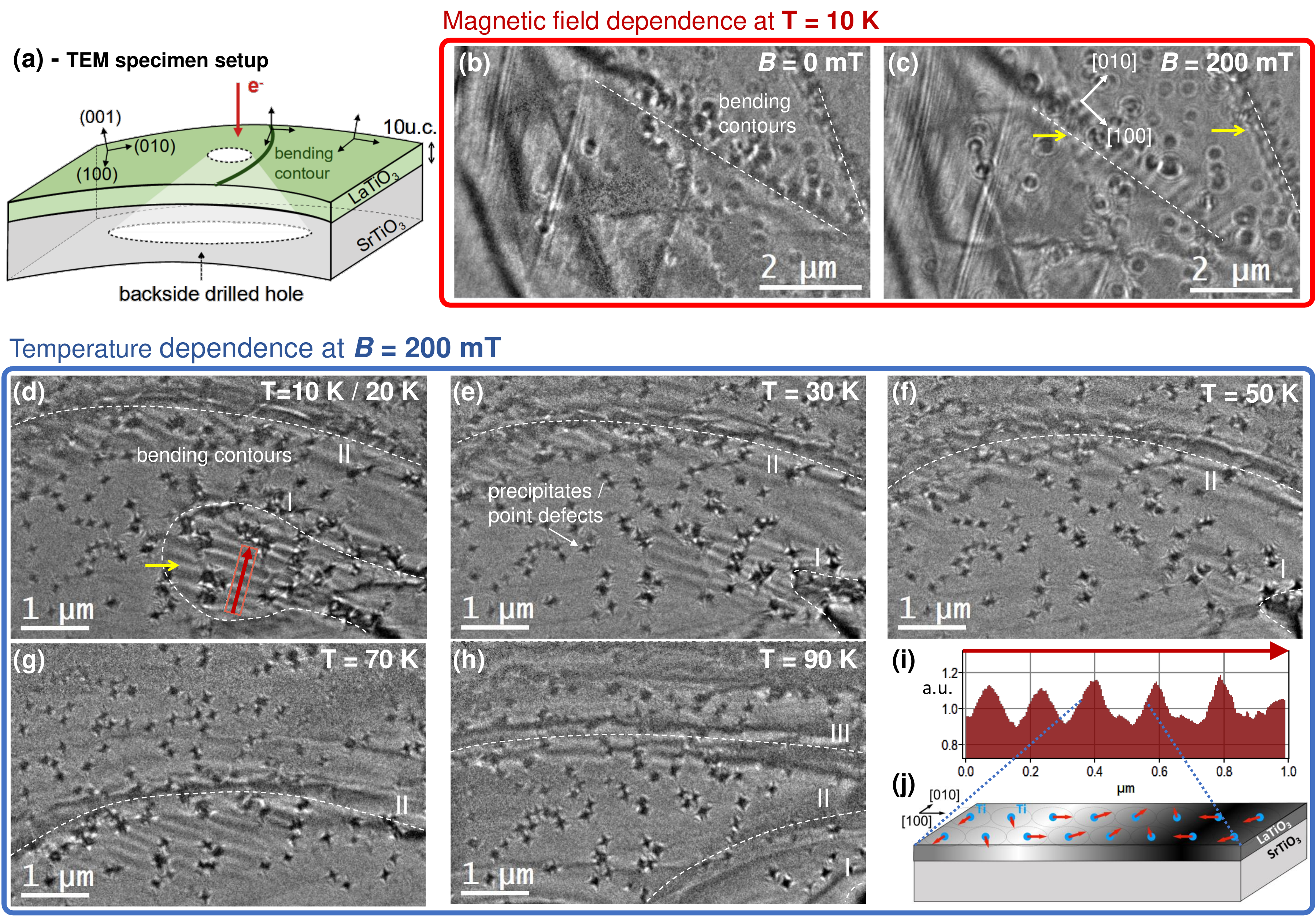}
	\caption{\textbf{ Temperature- and field-dependent Lorentz transmission electron microscopy (LTEM) of characteristic uniaxial magnetic modulation / texture.} \textbf{(a)} Scattering and TEM specimen geometry employed for LTEM.
	\textbf{(b)},\textbf{(c)} show LTEM images at $10~\mathrm{K}$ without magnetic field and with an applied out-of-plane field of $200~\mathrm{mT}$. The latter exhibits a magnetic texture (highlighted by yellow arrows) at bending contours, where the direction of modulation agrees with a $\langle 100 \rangle$ crystallographic axis of the cubic \ch{SrTiO3} substrate.
	\textbf{(d)}-\textbf{(h)} show the gradual shrinkage and motion of the magnetic texture associated with bending contours I and II with increasing temperature until $90$~K.
	\textbf{(i)} displays a line profile along the line scan over the uniaxially modulated region indicated as red arrow in \textbf{(d)} revealing a periodicity of 
	about $200~$nm.
	\textbf{(j)} Schematic depiction of the LTEM contrast (one period) by the spin modulation of the Ti atoms.}
	\label{Figure3}
\end{figure}

In order to search for relevant nanometer scale inhomogeneities in the temperature regime where the LCMR is observed, we carried out cryogenic Lorentz transmission electron microscopy (LTEM) on 10~u.c. thick \ch{LaTiO3} films (see Figure ~\ref{Figure3}(a) and methods). To reveal the nature of the observed contrast variations, we investigated their dependence on temperature, external magnetic field and defocus at a dedicated continuous-flow liquid Helium-cooled TEM instrument \cite{Boerrnert2019}.

At low temperature, under zero magnetic field only small dark dots distributed across the entire field of view are observed (see Figure ~\ref{Figure3}(b)). Those are invariant with temperature and field and may be associated to impurities, precipitates or point defects. Under external magnetic field, a stripe-like uniaxial modulation of the LTEM contrast is however visible at out-of-focus conditions. This stripe pattern is observed close to bending contours (i.e., loci of locally fulfilled Bragg conditions, which naturally occur in thin strained TEM lamellas, see Figure ~\ref{Figure3}(a),(c)). The modulation has a periodicity of $188\pm16$nm within the resolution of our measurement (Figure ~\ref{Figure3}(i)). Correlated electron diffraction data further reveals that the modulation direction aligns with the $\langle 010 \rangle$ axis of the underlying \ch{SrTiO3} sample (corresponding to the $\langle 110 \rangle$ orientation of the orthorhombic \ch{LaTiO3} film), i.e. along the Ti-O-Ti direction (Figure ~\ref{Figure3}(j) and Suppl. Info. IV). The observed contrast modulation is strongly field dependent and virtually vanishes at zero external field (Figure ~\ref{Figure3}(b),(c)). Moreover, the contrast vanishes in focus and changes sign upon inverting the sign of the focus (see Suppl. Info. IV). These observations are compatible with the presence of a uniaxial magnetic texture in the \ch{LaTiO3} films (see Suppl. Info. IV for further discussion of the contrast mechanism). To our knowledge, the existence of such a magnetic structure has not been previously reported in \ch{LaTiO3}.

To investigate the stability of the magnetic stripe texture, we performed temperature dependent measurements. The stripes appear stable between 8~K and 20~K. Above 20~K they start to disappear, where the shrinking of the modulated regions may be inhomogeneous. For instance, modulation associated to bending contour I seems to be gone at around 35K (not shown), whereas that associated to contour II persists up to 80K. We observe no clear temperature dependence of the periodicity of the modulation.

Owing to the unique contrast amplification mechanism, which makes the stripes only visible around the bending contours (see Suppl. Info. IV), we cannot determine the fractional coverage of the modulation. However, the shifting of the modulation with bending contour II in Figs.~\ref{Figure3}(d-g) indicates a very large coverage of several tens of percent.

\section{Discussion}
The uniaxial magnetic texture in the LTO film observed in LTEM covers a sizable area of the sample with a morphology that evolves with  temperature above 20~K up until 80~K when the pattern disappears. These features mirror those ascribed to the low mobility regions in magnetotransport. In the following we discuss a possible origin of the magnetic texture as well as its impact on magnetotransport.

Magnetic stripes were reported in STO-based heterostructures, originating from ferroelastic domain walls in STO substrates \cite{Christensen2019,Kalisky2013}. These domains however have typical sizes ranging from 5 to 15~$\mu$m \cite{Christensen2019,Kalisky2013}, which is more than one order of magnitude larger than those we observe. This casts some doubt on their role to explain our modulation. Experimental \cite{Meijer1999} and theoretical \cite{Schmitz2005} investigations established that bulk LaTiO$_{3}$ orders antiferromagnetically (AF) at low temperature with canted Ti$^{3+}$ moments owing to orthorhombic distortions, resulting in a small ferromagnetic component. Another possible explanation for the uniaxial modulation would then be ferromagnetic stripes with out-of-plane anisotropy. However, this scenario would imply out-of-plane magnetic domains separated by Bloch-like domain boundaries \cite{Gabay1985}, both of which would generate no contrast in our LTEM configuration.

Based on the large Rashba spin-orbit coupling of the LTO thin film \cite{Veit}, we propose an alternative scenario where a spiral magnetic phase is stabilized by the double exchange process between mobile carriers and static moments, owing to a large Hund's coupling energy \cite{Banerjee2013, Gabay2013}. According to the double exchange scenario, the spins $\mathbf{S}_{\mathbf{\vec{r}}}$ of the mobile carriers are locked to the local Ti$^{3+}$ magnetic moments. The ground state is determined by a balance between an effective exchange term $J \sim t n_f$ (with $t$ the hopping energy and $n_f$ the mobile electronic fraction per Ti atom), a Dzyaloshinskii-Moriya (DM) term $D \sim \alpha n_f$ (where $\alpha$ is the interfacial Rashba-like spin-orbit energy) and an exchange anisotropy term $A$\cite{Banerjee2013, Gabay2013}. The corresponding effective spin Hamiltonian reads:
\begin{equation}
\label{Ham2}
\begin{aligned}
    \cal{H}_{\textrm{eff }} = & \sum_{\mathbf{\vec{r}}} 
     -J(\mathbf{S}_{\mathbf{\vec{r}}}\mathbf{\cdot} \mathbf{S}_{\mathbf{\vec{r}+\hat{x}}a_0}+\mathbf{S}_{\mathbf{\vec{r}}}\mathbf{\cdot} \mathbf{S}_{\mathbf{\vec{r}+\hat{y}}a_0})-D\mathbf{\hat{z}}\mathbf{\cdot} \mathbf{S}_{\mathbf{\vec{r}}}\mathbf{\times} \mathbf{S}_{\mathbf{\vec{r}+\hat{Q}}a_0}\\
   &-A{S^\textrm{y}}_{\mathbf{\vec{r}}} {S^\textrm{y}}_{\mathbf{\vec{r}+\hat{x}}a_0} -A{S^\textrm{x}}_{\mathbf{\vec{r}}} {S^\textrm{x}}_{\mathbf{\vec{r}+\hat{y}}a_0}
 \end{aligned}
\end{equation}

where $\mathbf{\hat{z}}$ corresponds to the out-of-plane [001] direction, $\mathbf{a_0}$ is the in-plane Ti-O-Ti distance, and $\mathbf{\hat{x}}$ and $\mathbf{\hat{y}}$ denote the Ti-O-Ti bond directions (see Supplementary Material for further details). For the set of parameters relevant to the LTO/STO interface ($t \sim 250$ meV \cite{Ishida2008}, $\alpha \sim 2$ meV -- \cite{Biscaras2012}), the ground state configuration is a spiral, propagating with a wavevector $Q = \frac{2D}{J \textrm{a}_0}$ parallel to the Ti-O-Ti bonds ($\langle 100 \rangle$ or $\langle 010 \rangle$ direction in the pseudocubic LTO frame) and polarized in a plane parallel to the interface \cite{Banerjee2013, Gabay2013}. Here, in constrast to the \ch{LaAlO3}/\ch{SrTiO3} interface case when the DM term is in-plane, octahedral rotations in LTO/STO allow an out-of-plane DM contribution. This is important in our case since only an in-plane spiral (like the one shown in Figure ~\ref{Figure3}(j)) would result in a LTEM contrast.
We calculate the pitch of the spiral to be $\mathbf{d}=\frac{2\pi}{Q} \sim 160$~nm. Both the pitch and the modulation direction are in good agreement with our LTEM measurements. 

\begin{figure}
	\includegraphics[width=0.5\textwidth]{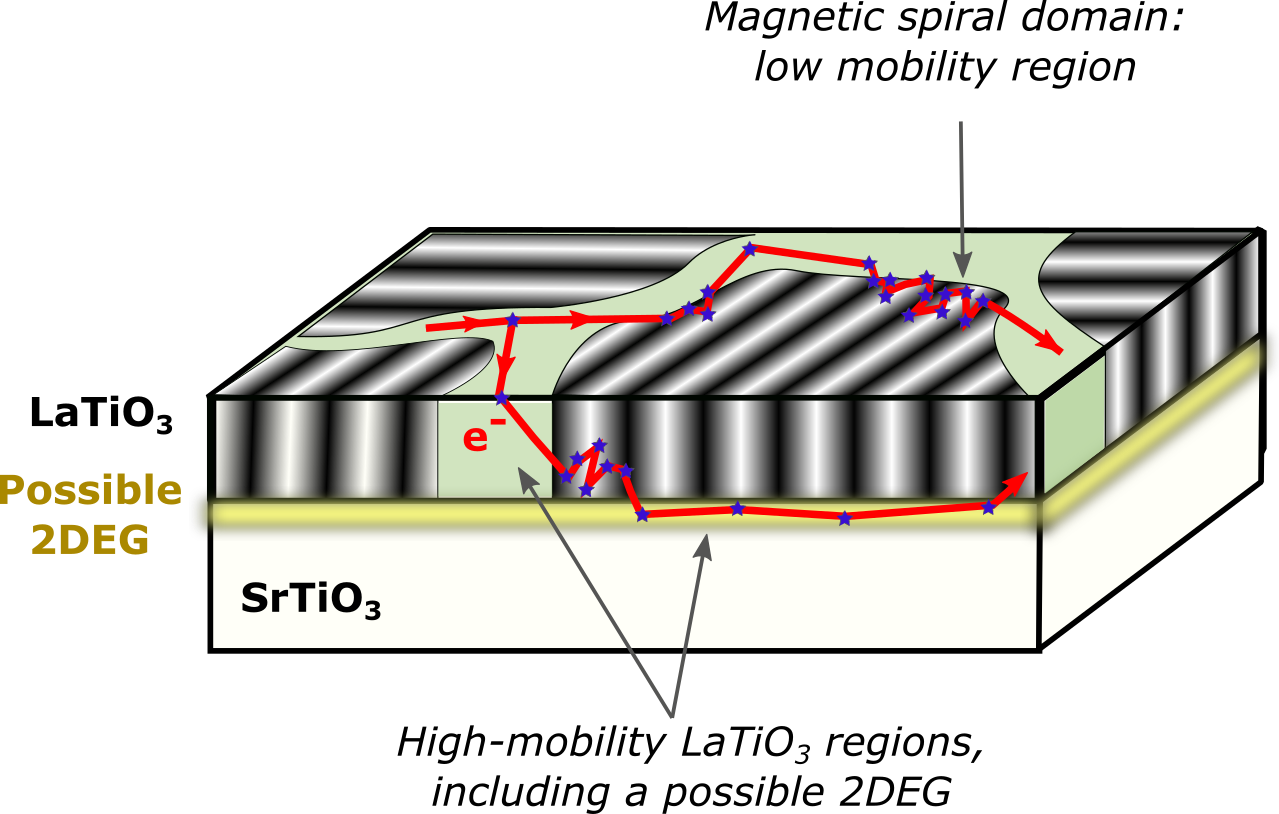}
	\caption{Schematic view of electronic transport in magnetic LTO thin films in the Parish-Littlewood framework. Spiral magnetic domains (represented by stripes) are associated with increased magnetic scattering, leading to lower mobility. In LTO regions which are not in the spiral ground state (light green), or a possible 2DEG (yellow), magnetic scattering is reduced, leading to higher mobility.}
	\label{Figure4}
\end{figure}

The existence of the magnetic spiral state in LTO strongly influence magnetotransport. In particular, in the magnetic region, magnetic scattering can significantly increase the scattering rate. Considering that a spin-spiral texture consist of successive interfaces between magnetic slabs with small tilt of the magnetisation between them, we estimate that the scattering rate in the spin-spiral domains is more than one order of magnitude larger than in the rest of the sample (see Suppl. Info. V). In the framework of the Parish-Littlewood scenario\cite{Parish1,Kozlova}, we propose that the low-mobility regions could correspond to the magnetic regions of the LTO film, where electrons would experience spin-scattering (see Figure ~\ref{Figure4}). This would naturally explain the high calculated coverage. The high-mobility channels could be comprised of either non-magnetic LTO regions, of a confined 2DEG or unintentionally doped STO layer, where the high-mobility would not be impacted as much by magnetic scattering. Since the proportion of LTO in the total conducting volume only increase with film thickness, the low-mobility region coverage would also increase with thickness, as is observed. The exact influence of a three-dimensional electron motion on the Parish-Littlewood theory is unclear, and should be further examined theoretically.
\\

\section{Conclusion:} To summarise our findings, \ch{LaTiO3} thin films on \ch{SrTiO3} were investigated using magnetotransport and transmission electron microscopy measurements. The main result is the observation of a non-saturating linear magnetoresistance, with a colossal amplitude up to 6500\% at 9T. This very large effect is understood as resulting from the combination of high carrier mobility (40 000 cm$^2$/V.s) with an extreme coverage of guiding-centre regions of 49\% up to 89\%. These lower-mobility regions are tied to a striped pattern of magnetic origin in the \ch{LaTiO3} film revealed by Lorentz TEM. The observed striped pattern is shown to be compatible with spiral magnetism.
Our study establishes a link between non-saturating linear colossal magnetoresistance and complex magnetic structure. This result suggests a new design concept for devices with large linear magnetoresistance in magnetic materials and heterostructures. We further expect it to trigger further theoretical interest on linear magnetoresistance in inhomogeneous and magnetic systems.
 


\section{Experimental Section}
\textbf{Growth}\\
For growth a laser flux of 1.5\,J.cm$^{-2}$ and a polycrystalline La$_2$Ti$_2$O$_7$ target at a distance of 55\,mm from the substrate were employed. Thin films with different thicknesses of 3, 5 and 10 unit cells (u.c.) were grown into Hall bars, with lengths of 20, 40, 60 and 80\,$\mu$m and widths of 50, 100 and 150\,$\mu$m. During growth, the band filling of LTO can be tuned by excess oxygen doping \cite{Scheiderer}. All samples investigated were grown under high oxygen background pressure ($1\times10^{-6}$\,mbar), inducing over oxidation and thus producing a correlated metallic system. Further details can be found in the supplementary information section 1. The samples were glued to chip carriers and wire bonded for the magnetotransport measurements.

\textbf{Transport measurements}\\
Magnetotransport experiments were performed using a Quantum Design physical properties measurement system (PPMS) in pseudo-AC mode, and with a ${}^3\mathrm{He}$-${}^4\mathrm{He}$ dilution refrigerator with the magnetic field applied perpendicular to the films, using standard lock-in techniques. The transverse magnetoresistance (MR) was studied with magnetic fields up to 11T, generated by a superconducting magnet. In Figures 1 and 2, several Hallbars of different sizes are investigated: Hallbars 1 are 20x50\,$\mu$m, Hallbars 2 are 20x80\,$\mu$m, and Hallbars 3 are 50x150\,$\mu$m.

\textbf{TEM preparation and measurements}\\
Electron-transparent TEM samples have been prepared by a classical back-side thinning procedure consisting of (A) manual coarse mechanical grinding, (B) dimpling, and (C) final Ar ion milling using a precision ion polishing system (PIPS 691, Gatan Inc., US) until a hole appears. All thinning steps have been applied from the back (substrate) side of the sample in order to preserve the \ch{LaTiO3} film. The latter was verified by employing TEM-EDX, which showed the presence of La throughout the entire sample region of interest. Cryogenic TEM investigations were carried out at the JEOL 2010F Dresden special instrument \cite{Boerrnert2019} operated at an acceleration voltage of 200~kV. The microscope is fitted with a custom-made continuous-flow liquid Helium cryostat enabling stable cooling for several days while varying temperature, which is essential for the above studies. External magnetic fields were applied by exciting the objective lens coils of the TEM. Lorentz imaging (i.e., out-of-focus and small external field conditions) was set by varying the first transfer lens (TL11) of the CETCOR imaging corrector utilized as a pseudo Lorentz lens. The defocus in the field series (Figure ~3) as well as in the focal series (Suppl. Figure  S6) was adjusted by inspecting the diffraction (Fresnel fringes) at the sample edge. The background of the TEM images in Figure 3(b-h) due to thickness variations of the wedge-shaped sample (see Figure 3(a)) was subtracted to enhance the magnetic contrast.

\medskip
\textbf{Acknowledgements} \par 
The authors are grateful to S. Kuhn for expert assistance in sample processing. 
L.V. was supported by the Leibniz Association through the Leibniz Competition. A.L. acknowledges funding from the European Research Council (ERC) under the Horizon 2020 research and innovation program of the European Union (grant agreement number 715620). 
This work was supported by the Deutsche Forschungsgemeinschaft (DFG) through projects 461150024 and 431448015, through the Wuerzburg-Dresden Cluster of Excellence on Complexity and Topology in Quantum Matter - ct.qmat (EXC 2147, project-id 390858490) and the Collaborative Research Center SFB 1170 "ToCoTronics" (project-id 258499086).

\medskip

\nocite{EndofMaintext}

\bibliographystyle{unsrt}
\bibliography{LaTiO3paper_v8}{}

\begin{thebibliography}{10}

\bibitem{Abrikosov98}
A.~A. Abrikosov.
\newblock Quantum magnetoresistance.
\newblock {\em Phys. Rev. B}, 58:2788--2794, Aug 1998.

\bibitem{Majumder}
D.~D. Majumder and S.~Karan.
\newblock {\em Magnetic properties of ceramic nanocomposites}, pages 51--91.
\newblock 12 2013.

\bibitem{Jin1994}
S.~Jin, T.~H. Tiefel, M.~McCormack, R.~A. Fastnacht, R.~Ramesh, and L.~H. Chen.
\newblock Thousandfold change in resistivity in magnetoresistive
  $\mathrm{LaCaMnO}$ films.
\newblock {\em Science}, 264(5157):413--415, 1994.

\bibitem{Lin2016}
C.~Lin, C.~Yi, Y.~Shi, L.~Zhang, G.~Zhang, J.~M\"uller, and Y.~Li.
\newblock Spin correlations and colossal magnetoresistance in
  $\mathrm{{HgCr}_{2}{Se}_{4}}$.
\newblock {\em Phys. Rev. B}, 94(22), December 2016.

\bibitem{Sinjukow2003}
P.~Sinjukow and W.~Nolting.
\newblock Metal-insulator transition in $\mathrm{EuO}$.
\newblock {\em Phys. Rev. B}, 68(12), September 2003.

\bibitem{Jin}
S.~Jin, M.~McCormack, T.~H. Tiefel, and R.~Ramesh.
\newblock Colossal magnetoresistance in $\mathrm{{La}{Ca}{Mn}{O}}$
  ferromagnetic thin films.
\newblock {\em Journal of Applied Physics}, 76(10):6929--6933, 1994.

\bibitem{Subramanian}
M.~A. Subramanian, B.~H. Toby, A.~P. Ramirez, W.~J. Marshall, A.~W. Sleight,
  and G.~H. Kwei.
\newblock Colossal magnetoresistance without $\mathrm{{Mn}^{3+}/{Mn}^{4+}}$
  double exchange in the stoichiometric pyrochlore
  $\mathrm{{Tl}_{2}{Mn}_{2}{O}_{7}}$.
\newblock {\em Science}, 273(5271):81--84, 1996.

\bibitem{Roeder}
H.~R\"oder, J.~Zang, and A.~R. Bishop.
\newblock Lattice effects in the colossal-magnetoresistance manganites.
\newblock {\em Phys. Rev. Lett.}, 76:1356--1359, Feb 1996.

\bibitem{Maksimovic2020}
N.~Maksimovic, I.~M. Hayes, V.~Nagarajan, J.~G. Analytis, A.~E. Koshelev,
  J.~Singleton, Y.~Lee, and T.~Schenkel.
\newblock {Magnetoresistance Scaling and the Origin of $H$-Linear Resistivity
  in
  ${\mathrm{BaFe}}_{2}({\mathrm{As}}_{1\ensuremath{-}x}{\mathrm{P}}_{x}{)}_{2}$}.
\newblock {\em Phys. Rev. X}, 10(4), December 2020.

\bibitem{McGuire2021}
M.~A. McGuire, Q.~Zhang, H.~Miao, W.~Luo, M.~Yoon, Y.~Liu, T.~Yilmaz, and
  E.~Vescovo.
\newblock Antiferromagnetic order and linear magnetoresistance in
  fe-substituted shandite $\mathrm{{Co}_{3}{In}_{2}{S}_{2}}$.
\newblock {\em Chem. Mater.}, 2021.

\bibitem{Wei}
F.~Wei, X.~P.~A. Gao, S.~Ma, and Z.~Zhang.
\newblock Giant linear magnetoresistance and carrier density tunable transport
  in topological crystalline insulator $\mathrm{SnTe}$ thin film.
\newblock {\em physica status solidi (b)}, 256(10):1900139, 2019.

\bibitem{WangX}
X.~Wang, Y.~Du, S.~Dou, and C.~Zhang.
\newblock Room temperature giant and linear magnetoresistance in topological
  insulator $\mathrm{{Bi}_{2}{Te}_{3}}$ nanosheets.
\newblock {\em Phys. Rev. Lett.}, 108:266806, Jun 2012.

\bibitem{Zhang}
W.~Zhang, R.~Yu, W.~Feng, Y.~Yao, H.~Weng, X.~Dai, and Z.~Fang.
\newblock Topological aspect and quantum magnetoresistance of
  $\ensuremath{\beta}\mathrm{\text{\ensuremath{-}}}{\mathrm{ag}}_{2}\mathrm{Te}$.
\newblock {\em Phys. Rev. Lett.}, 106:156808, Apr 2011.

\bibitem{Shoenberg}
D.~Shoenberg.
\newblock {\em Magnetic Oscillations in Metals}.
\newblock Cambridge Univ. Press, 1984.

\bibitem{Datta}
S.~Datta.
\newblock {\em Electronic Transport in mesoscopic systems}.
\newblock Cambridge Univ. Press, 1997.

\bibitem{Parish1}
M.~M. Parish and P.~B. Littlewood.
\newblock Non-saturating magnetoresistance in heavily disordered
  semiconductors.
\newblock {\em Nature}, 426(6963):162--165, 2003.

\bibitem{Parish2}
M.~M. Parish and P.~B. Littlewood.
\newblock Classical magnetotransport of inhomogeneous conductors.
\newblock {\em Phys. Rev. B}, 72:094417, Sep 2005.

\bibitem{Kozlova}
N.~V. Kozlova, N.~Mori, O.~Makarovsky, L.~Eaves, Q.~D. Zhuang, A.~Krier, and
  A.~Patan{\`e}.
\newblock Linear magnetoresistance due to multiple-electron scattering by
  low-mobility islands in an inhomogeneous conductor.
\newblock {\em Nature Communications}, 3(1):1097, 2012.

\bibitem{Mallik2021}
S~Mallik, G.C. Menard, G.~Saiz, I.~Gilmutdinov, D.~Vignolles, C.~Proust,
  A.~Gloter, N.~Bergeal, M.~Gabay, and M~Bibes.
\newblock From low-field sondheimer oscillations to high-field very large and
  linear magnetoresistance in a $\mathrm{{SrTiO}_{3}}$-based two-dimensional
  electron gas.
\newblock {\em Nano Letters}, 2022(22):65--72, December 2021.

\bibitem{Song}
J.~C.~W. Song, G.~Refael, and P.~A. Lee.
\newblock Linear magnetoresistance in metals: Guiding center diffusion in a
  smooth random potential.
\newblock {\em Phys. Rev. B}, 92:180204, Nov 2015.

\bibitem{Meng2020}
J.~Meng, X.~Chen, M.~Liu, W.~Jiang, Z.~Zhang, J.~Ling, T.~Shao, C.~Yao, L.~He,
  R.~Dou, C.~Xiong, and J.~Nie.
\newblock Large linear magnetoresistance caused by disorder in
  $\mathrm{{WTe}_{2-\delta}}$ thin film.
\newblock {\em Journal of Physics: Condensed Matter}, 32(35):355703, jun 2020.

\bibitem{Yang}
X.~Yang, H.~Bai, Z.~Wang, Y.~Li, Q.~Chen, J.~Chen, Y.~Li, C.~Feng, Y.~Zheng,
  and Z.~Xu.
\newblock Giant linear magneto-resistance in nonmagnetic
  $\mathrm{{Pt}{Bi}_{2}}$.
\newblock {\em Applied Physics Letters}, 108(25):252401, 2016.

\bibitem{Xu}
R.~Xu, A.~Husmann, T.~F. Rosenbaum, M.~L. Saboungi, J.~E. Enderby, and P.~B.
  Littlewood.
\newblock Large magnetoresistance in non-magnetic silver chalcogenides.
\newblock {\em Nature}, 390(6655):57--60, 1997.

\bibitem{Husmann}
A.~Husmann, J.~B. Betts, G.~S. Boebinger, A.~Migliori, T.~F. Rosenbaum, and
  M.~L. Saboungi.
\newblock Megagauss sensors.
\newblock {\em Nature}, 417(6887):421--424, 2002.

\bibitem{Lee}
M.~Lee, T.~F. Rosenbaum, M.-L. Saboungi, and H.~S. Schnyders.
\newblock Band-gap tuning and linear magnetoresistance in the silver
  chalcogenides.
\newblock {\em Phys. Rev. Lett.}, 88:066602, Jan 2002.

\bibitem{Novak}
M.~Novak, S.~Sasaki, K.~Segawa, and Y.~Ando.
\newblock Large linear magnetoresistance in the dirac semimetal
  $\mathrm{TlBiSSe}$.
\newblock {\em Phys. Rev. B}, 91:041203, Jan 2015.

\bibitem{Huang}
X.~Huang, L.~Zhao, Y.~Long, P.~Wang, D.~Chen, Z.~Yang, H.~Liang, M.~Xue,
  H.~Weng, Z.~Fang, X.~Dai, and G.~Chen.
\newblock Observation of the chiral-anomaly-induced negative magnetoresistance
  in 3d weyl semimetal $\mathrm{{Ta}{As}}$.
\newblock {\em Phys. Rev. X}, 5:031023, Aug 2015.

\bibitem{Xiong}
J.~Xiong, S.~Kushwaha, J.~Krizan, T.~Liang, R.~J. Cava, and N.~P. Ong.
\newblock Anomalous conductivity tensor in the dirac semimetal
  $\mathrm{{Na}_{2}{Bi}}$.
\newblock {\em {EPL} (Europhysics Letters)}, 114(2):27002, apr 2016.

\bibitem{Narayanan}
A.~Narayanan, M.~D. Watson, S.~F. Blake, N.~Bruyant, L.~Drigo, Y.~L. Chen,
  D.~Prabhakaran, B.~Yan, C.~Felser, T.~Kong, P.~C. Canfield, and A.~I. Coldea.
\newblock Linear magnetoresistance caused by mobility fluctuations in $n$-doped
  $\mathrm{{Cd}_{3}{As}_{2}}$.
\newblock {\em Phys. Rev. Lett.}, 114:117201, Mar 2015.

\bibitem{Feng}
J.~Feng, Y.~Pang, D.~Wu, Z.~Wang, H.~Weng, J.~Li, X.~Dai, Z.~Fang, Y.~Shi, and
  L.~Lu.
\newblock Large linear magnetoresistance in dirac semimetal
  $\mathrm{{Cd}_{3}{As}_{2}}$ with fermi surfaces close to the dirac points.
\newblock {\em Phys. Rev. B}, 92:081306, Aug 2015.

\bibitem{Zhang2021}
J.~Zhang, J.~M. Ok, Y.-Y. Pai, J.~Lapano, E.~Skoropata, A.~R. Mazza, H.~Li,
  A.~Huon, S.~Yoon, B.~Lawrie, M.~Brahlek, T.~Z. Ward, G.~Eres, H.~Miao, and
  H.~N. Lee.
\newblock Extremely large magnetoresistance in high-mobility
  $\mathrm{{SrNbO}_{3}/{SrTiO}_{3}}$ heterostructures.
\newblock {\em Phys. Rev. B}, 104(16), October 2021.

\bibitem{Abrikosov}
A.~A. Abrikosov.
\newblock Quantum linear magnetoresistance.
\newblock {\em Europhysics Letters ({EPL})}, 49(6):789--793, mar 2000.

\bibitem{He2021}
Y.~He, J.~Gayles, M.~Yao, T.~Helm, T.~Reimann, V.~Strocov, W.~Schnelle,
  M.~Nicklas, Y.~Sun, G.~H. Fecher, and C.~Felser.
\newblock Large linear non-saturating magnetoresistance and high mobility in
  ferromagnetic $\mathrm{MnBi}$.
\newblock {\em Nat. Comm.}, 2021.

\bibitem{Niu2021}
W.~Niu, Y.~Gan, Z.~Wu, X.~Zhang, Y.~Wang, Y.~Chen, L.~Wang, Y.~Xu, L.~He,
  Y.~Pu, and X.~Wang.
\newblock {Large Linear Magnetoresistance of High-Mobility 2D Electron System
  at Nonisostructural $\gamma$-$\mathrm{{Al2O}_{3}/{SrTiO}_{3}}$
  Heterointerfaces}.
\newblock {\em Advanced Materials Interfaces}, 8(21):2101235, 2021.

\bibitem{Wang2021}
Z.-C. Wang, L.~Chen, S.-S. Li, J.-S. Ying, F.~Tang, G.-Y. Gao, Y.~Fang,
  W.~Zhao, D.~Cortie, X.~Wang, and R.-K. Zheng.
\newblock Giant linear magnetoresistance in half-metallic
  $\mathrm{{Sr}_{2}{Cr}{Mo}{O}_{6}}$ thin films.
\newblock {\em npj Quantum Materials}, 6(1):53, 2021.

\bibitem{Schneider2006}
C.~W. Schneider, S.~Thiel, G.~Hammerl, C.~Richter, and J.~Mannhart.
\newblock Microlithography of electron gases formed at interfaces in oxide
  heterostructures.
\newblock {\em Applied Physics Letters}, 89(12):122101, 2006.

\bibitem{Biscaras}
J.~Biscaras, N.~Bergeal, A.~Kushwaha, T.~Wolf, A.~Rastogi, R.~C. Budhani, and
  J.~Lesueur.
\newblock Two-dimensional superconductivity at a mott insulator/band insulator
  interface $\mathrm{{LaTiO}_{3}/{SrTiO}_{3}}$.
\newblock {\em Nature Communications}, 1(1):89, 2010.

\bibitem{Lebedev2020}
N.~Lebedev, M.~Stehno, A.~Rana, N.~Gauquelin, J.~Verbeeck, A.~Brinkman, and
  J.~Aarts.
\newblock Inhomogeneous superconductivity and quasilinear magnetoresistance at
  amorphous $\mathrm{{LaTiO}_{3}/{SrTiO}_{3}}$ interfaces.
\newblock {\em Journal of Physics: Condensed Matter}, 33(5):055001, feb 2020.

\bibitem{Scheiderer}
P.~Scheiderer, M.~Schmitt, J.~Gabel, M.~Zapf, M.~St{\"u}binger, P.~Sch{\"u}tz,
  L.~Dudy, C.~Schlueter, T.-L. Lee, M.~Sing, and R.~Claessen.
\newblock Tailoring materials for mottronics: Excess oxygen doping of a
  prototypical mott insulator.
\newblock {\em Advanced Materials}, 30(25):1706708, 2018.

\bibitem{Ramirez}
A.~P. Ramirez.
\newblock Colossal magnetoresistance.
\newblock {\em Journal of Physics: Condensed Matter}, 9(39):8171--8199, sep
  1997.

\bibitem{Veit}
M.~J. Veit, R.~Arras, B.~J. Ramshaw, R.~Pentcheva, and Y.~Suzuki.
\newblock Nonzero berry phase in quantum oscillations from giant rashba-type
  spin splitting in $\mathrm{{LaTiO}_{3}/{SrTiO}_{3}}$ heterostructures.
\newblock {\em Nature Communications}, 9(1):1458, 2018.

\bibitem{Dufouleur2016}
J.~Dufouleur, L.~Veyrat, B.~Dassonneville, C.~Nowka, S.~Hampel, P.~Leksin,
  B.~Eichler, O.~G. Schmidt, B.~B\"uchner, and R.~Giraud.
\newblock Enhanced mobility of spin-helical dirac fermions in disordered 3d
  topological insulators.
\newblock {\em Nano Letters}, 16(11):6733--6737, 2016.
\newblock PMID: 27706936.

\bibitem{Lonchakov2021}
A.~T. Lonchakov and S.~B. Bobin.
\newblock Large linear magnetoresistance in single $\mathrm{HgSe}$ crystals
  induced by low-concentration co impurity.
\newblock {\em Applied Physics Letters}, 118(6):062106, 2021.

\bibitem{Jin2016}
H.~Jin, K.~Lee, S.~H. Baek, J.~S. Kim, B.~K. Cheong, B.~H. Park, S.~Yoon, B.~J.
  Suh, C.~Kim, S.~S. Seo, and S.~Lee.
\newblock Large linear magnetoresistance in heavily-doped
  $\mathrm{{Nb}:{SrTiO}_{3}}$ epitaxial thin films.
\newblock {\em Sci Rep.}, October 2016.

\bibitem{Kormondy2018}
K.~J. Kormondy, L.~Gao, X.~Li, S.~Lu, A.~B. Posadas, S.~Shen, M.~Tsoi, M.~R.
  McCartney, D.~J. Smith, J.~Zhou, L.~L. Lev, M.~A. Husanu, V.~N. Strocov, and
  A.~A. Demkov.
\newblock Large positive linear magnetoresistance in the two-dimensional t 2g
  electron gas at the $\mathrm{{EuO}/{SrTiO}_{3}}$ interface.
\newblock {\em Sci Rep.}, May 2018.

\bibitem{Hu}
J.~Hu, M.~M. Parish, and T.~F. Rosenbaum.
\newblock Nonsaturating magnetoresistance of inhomogeneous conductors:
  Comparison of experiment and simulation.
\newblock {\em Phys. Rev. B}, 75:214203, Jun 2007.

\bibitem{J.Wang}
W.~J. Wang, K.~H. Gao, Z.~Q. Li, T.~Lin, J.~Li, C.~Yu, and Z.~H. Feng.
\newblock Classical linear magnetoresistance in epitaxial graphene on
  $\mathrm{SiC}$.
\newblock {\em Applied Physics Letters}, 105(18):182102, 2014.

\bibitem{Patane}
A.~Patan{\`e}, A.~Feu, W.~Makarovsky, O.~Drachenko, O.~Eaves, L.~Krier,
  A.~Zhuang, Q.~Helm, M.~Goiran, M.~Hill, and G.
\newblock Effect of low nitrogen concentrations on the electronic properties of
  $\mathrm{{InAs}_{1\ensuremath{-}x}{N}_{x}}$.
\newblock {\em Phys. Rev. B}, 80:115207, Sep 2009.

\bibitem{Boerrnert2019}
F.~Boerrnert, F.~Kern, F.~Harder, T.~Riedel, H.~Mueller, B.~B\"uchner, and
  A.~Lubk.
\newblock {The Dresden in-situ (S)TEM special with a continuous-flow
  liquid-helium cryostat}.
\newblock {\em {Ultramicroscopy}}, {203}:{12--20}, {2019}.

\bibitem{Christensen2019}
D~V Christensen, Y~Frenkel, Y~Z Chen, Y~W Xie, Z~Y Chen, Y~Hikita, A~Smith,
  L~Klein, H~Y Hwang, N~Pryds, and B~Kalisky.
\newblock {Strain-tunable magnetism at oxide domain walls}.
\newblock {\em Nature Physics}, 15(3):269--274, 2019.

\bibitem{Kalisky2013}
B.~Kalisky, E.~M Spanton, H.~Noad, J.~R Kirtley, K.~C Nowack, C.~Bell, H.~K
  Sato, M.~Hosoda, Y.~Xie, Y.~Hikita, G.~Woltmann, C.and~Pfanzelt, R.~Jany,
  C.~Richter, H.~Y Hwang, J.~Mannhart, and K.~A Moler.
\newblock {Locally enhanced conductivity due to the tetragonal domain structure
  in $\mathrm{{LaAlO}_{3}/{SrTiO}_{3}}$ heterointerfaces}.
\newblock {\em Nature Materials}, 12(12):1091--1095, 2013.

\bibitem{Meijer1999}
G.~I. Meijer, W.~Henggeler, J.~Brown, O.-S. Becker, J.~G. Bednorz, C.~Rossel,
  and P.~Wachter.
\newblock Reduction of ordered moment in strongly correlated
  ${\mathrm{latio}}_{3+\ensuremath{\delta}}$ upon band filling.
\newblock {\em Phys. Rev. B}, 59:11832--11836, May 1999.

\bibitem{Schmitz2005}
R.~Schmitz, O.~Entin-Wohlman, A.~Aharony, A.~B. Harris, and
  E.~M\"uller-Hartmann.
\newblock Magnetic structure of the jahn-teller system
  $\mathrm{{La}{Ti}{O}_{3}}$.
\newblock {\em Phys. Rev. B}, 71:144412, Apr 2005.

\bibitem{Gabay1985}
M.~Gabay and T.~Garel.
\newblock Phase transitions and size effects in the ising dipolar magnet.
\newblock {\em J. Physique}, 46:5--16, 1985.

\bibitem{Banerjee2013}
S.~Banerjee, O.~Erten, and M.~Randeria.
\newblock {Ferromagnetic exchange, spin-orbit coupling and spiral magnetism at
  the $\mathrm{{LaAlO}_{3}/{SrTiO}_{3}}$ interface}.
\newblock {\em Nat. Phys.}, 9:626--630, 2013.

\bibitem{Gabay2013}
M.~Gabay and J-M. Triscone.
\newblock {Hund rules with a twist}.
\newblock {\em Nat. Phys.}, 9:610--611, 2013.

\bibitem{Ishida2008}
H.~Ishida and A.~Liebsch.
\newblock {Origin of metallicity of $\mathrm{{LaTiO}_{3}/{SrTiO}_{3}}$
  heterostructures}.
\newblock {\em Phys. Rev. B}, 77:115350, 2008.

\bibitem{Biscaras2012}
J.~Biscaras.
\newblock {\em Two-Dimensional Superconductivity at Titanium Oxide Interfaces.}
\newblock Thesis, chapter 5, {Universit{\'e} Pierre et Marie Curie - Paris VI},
  December 2012.

\bibitem{EndofMaintext}
\textcolor{red}{Here ends the references from the main text.}

\bibitem{tanuma_calculation_2003}
S.~Tanuma, C.~J. Powell, and D.~R. Penn.
\newblock Calculation of electron inelastic mean free paths ({IMFPs}) {VII}.
  {Reliability} of the {TPP}-{2M} {IMFP} predictive equation.
\newblock {\em Surface and Interface Analysis}, 35(3):268--275, 2003.
\newblock \_eprint: https://onlinelibrary.wiley.com/doi/pdf/10.1002/sia.1526.

\bibitem{ohtomo_artificial_2002}
A.~Ohtomo, D.~A. Muller, J.~L. Grazul, and H.~Y. Hwang.
\newblock Artificial charge-modulationin atomic-scale perovskite titanate
  superlattices.
\newblock {\em Nature}, 419(6905):378--380, September 2002.
\newblock Number: 6905 Publisher: Nature Publishing Group.

\bibitem{Sondheimer}
E.~H. Sondheimer.
\newblock The influence of a transverse magnetic field on the conductivity of
  thin metallic films.
\newblock {\em Phys. Rev.}, 80:401--406, Nov 1950.

\bibitem{Harrison1960}
W.~A. Harrison.
\newblock {Electronic Structure of Polyvalent Metals}.
\newblock {\em Phys. Rev.}, 118(5):1190--1208, jun 1960.

\bibitem{vanDelft2021}
M.~R. van Delft, Y.~Wang, C.~Putzke, J.~Oswald, G.~Varnavides, C.~A~C Garcia,
  C.~Guo, H.~Schmid, V.~S{\"{u}}ss, H.~Borrmann, J.~Diaz, Y.~Sun, C.~Felser,
  B.~Gotsmann, P.~Narang, and P.~J.~W. Moll.
\newblock {Sondheimer oscillations as a probe of non-ohmic flow in WP2
  crystals}.
\newblock {\em Nature Communications}, 12(1):4799, 2021.

\bibitem{Lubk2018}
Axel Lubk.
\newblock Holography and tomography with electrons.
\newblock In {\em Advances in Imaging and Electron Physics}, volume 206. 2018.

\bibitem{Kurdestany2017}
J.~M. Kurdestany and S.~Satpathy.
\newblock Mott metal-insulator transition in the doped hubbard-holstein model.
\newblock {\em Phys. Rev. B}, 96:085132, Aug 2017.

\bibitem{Nakagawa2006}
N.~Nakagawa, Harold~Y. Hwang, and David~A. Muller.
\newblock Why some interfaces cannot be sharp.
\newblock {\em Nature Materials}, 5:204--209, 2006.

\bibitem{MacDonald2006}
A.~S. N\'u\~nez, R.~A. Duine, Paul Haney, and A.~H. MacDonald.
\newblock Theory of spin torques and giant magnetoresistance in
  antiferromagnetic metals.
\newblock {\em Phys. Rev. B}, 73:214426, Jun 2006.

\end{thebibliography}

\begin{figure}
\medskip
\captionsetup{labelformat=empty}
  \includegraphics[width=0.5\textwidth]{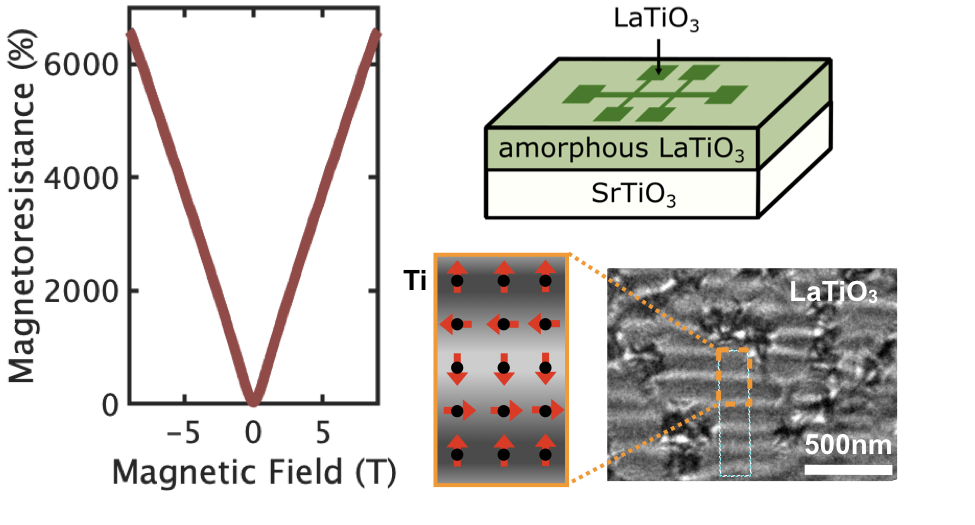}
  \medskip
  \caption{ \textbf{Graphical Table of Content} Linear magnetoresistance is of central importance for both fundamental research and technological application for its non-saturating character, especially for memory, electronics and sensors. Here linear magnetoresistance with colossal amplitude is observed in oxide thin films of LaTiO3 on SrTiO3. This very unusual effect stems from underlying magnetic spiral state, opening new avenue for devices with large linear magnetoresistance.}
\end{figure}


\clearpage

\setcounter{section}{0}
\setcounter{figure}{0}
\renewcommand{\thefigure}{S\arabic{figure}}

\section{Growth and spectroscopy}

\begin{figure*}[h!]
	\includegraphics[width=1\textwidth]{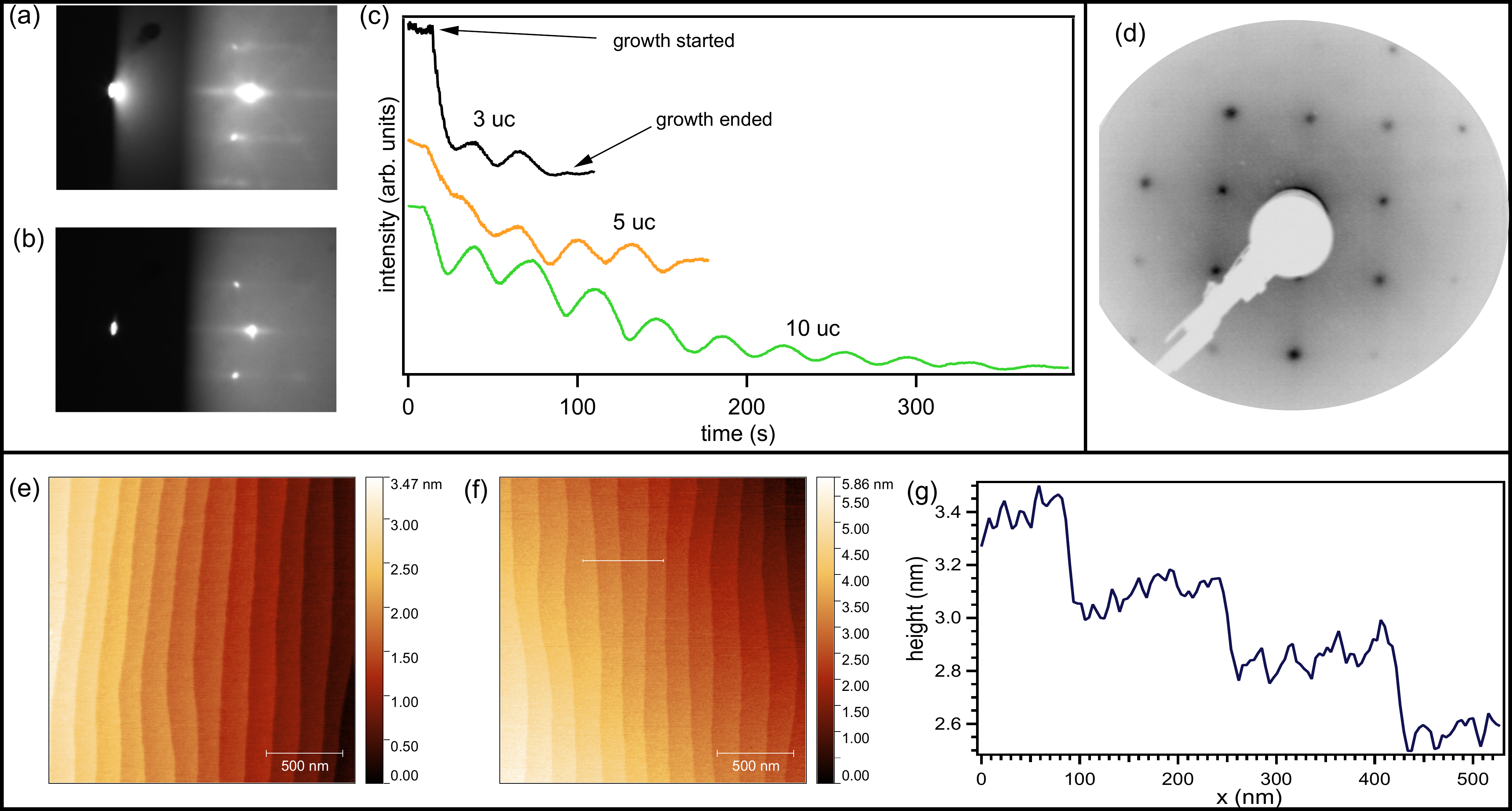}
	\caption{\textbf{Film growth and characterization}, \textbf{(a)} RHEED pattern of the bare substrate, \textbf{(b)}, RHEED pattern of the completed 5\,uc LaTiO$_3$ sample and \textbf{(c)}, RHEED intensity of the specular spot during growth for a 3\,uc, a 5\,uc and a 10\,uc sample. Each oscillation corresponds to a completed unit cell layer. \textbf{(d)}, LEED image of the film surface at a electron kinetic energy of 120\,eV. The clean 1\,$\times$\,1 reconstruction is imposed by the STO substrate. \textbf{(e)}, AFM image of the bare substrate. \textbf{(f)}, AFM image of a 5\,uc LaTiO$_3$ film, with \textbf{(g)}, a line profile. Large flat terraces with unit cell high steps of roughly 4\,\r{A} are visible.}
	\label{FigureS_growth}
\end{figure*}

The film thickness and quality is probed by reflection high energy electron diffraction (RHEED), exemplarily depicted for the 5\,u.c. LaTiO$_3$ (LTO) film from the main manuscript in Fig. \ref{FigureS_growth} (a-c). The diffraction pattern of the LTO film stays the same as the substrate indicating growth fully strained to the substrate. At room temperature, the lattice mismatch between LTO and \ch{SrTiO3} (STO) is 1.74\%, while below the tetragonal transition of STO it increases to 1.8\%. The intensity of the specular spot is monitored during growth to determine the film thickness. Each oscillation corresponds to a completed LTO layer. Due to the lithographic process, during which large areas of the bare substrate are covered with amorphous Al$_2$O$_3$, the following low energy electron diffraction (LEED) and photoemission spectroscopy data is taken on a non patterned reference sample, as the measurement area of these methods eclipses the pattern size significantly. The LEED image in Fig. \ref{FigureS_growth}d confirms the high crystal quality of the LTO film especially at the film surface. The 1\,$\times$\,1 diffraction pattern is imposed by the SrTiO$_3$ substrate.\\

\begin{figure*}[h]
	\includegraphics[width=1\textwidth]{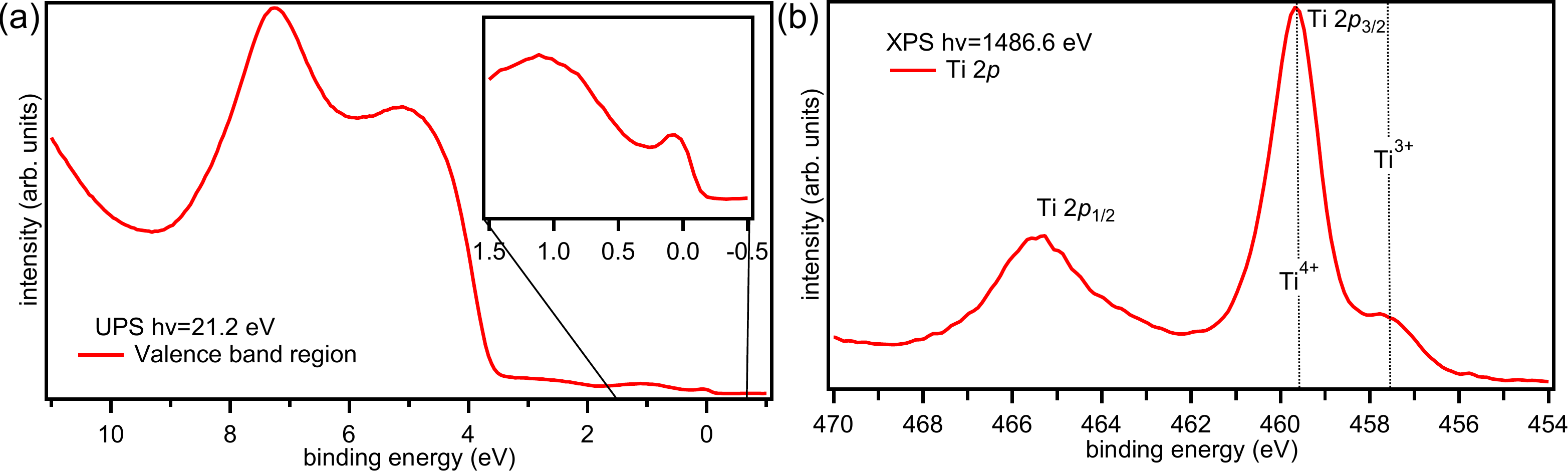}
	\caption{\textbf{Photoemission spectroscopy}, \textbf{(a)}, Ultraviolet photoemission spectra of the valence band region with zoom in on the Ti\,3\textit{d} containing region just below the Fermi energy. \textbf{(b)}, x-ray photoemission spectra of the Ti\,2\textit{p} core level.}
	\label{FigureS_PES}
\end{figure*}

The TiO$_2$ terminated substrates possess large flat terraces with unit cell high steps as can be inferred from the atomic force microscope image in Fig. \ref{FigureS_growth}(e). Another indicator for the high quality of the LTO films is, that their surfaces are barely discernible from the substrate in AFM, as demonstrated in the image of a 5\,u.c. film with line profile in Fig. \ref{FigureS_growth}(f,g).\\
By \textit{in-situ} transfer to a photoemission chamber the electronic state of the samples is independently investigated. Using ultra violet light (h$v=21.2$\,eV) we probe the valence band region as can be seen in Fig. \ref{FigureS_PES}(a) for a 3\,u.c. LTO sample. At the Fermi energy a quasiparticle peak and just 1.1 eV below it the lower Hubbard band are observed. The observation of the former in particular indicates metallic behavior. Due to the very limited inelastic mean free path of the photoelectrons at this excitation energy (approximately 0.1\,nm according to Tanuma \textit{et al.} \cite{tanuma_calculation_2003}), we can conclude, that the LTO film itself in addition to the interface \cite{ohtomo_artificial_2002} is metallic. The most likely explanation for that appears to be excess oxygen \textit{p}-doping \cite{Scheiderer}.\\

To check this hypothesis we analyze the Ti\,2\textit{p} core level by x-ray photoemission spectroscopy with monochromatized Al-K$_\alpha$ radiation (h$v$=1487\,eV). As can be seen in Fig. \ref{FigureS_PES}(b) for the same 3\,uc LTO sample from above, the spectra is dominated by the Ti$^{4+}$ valency, indicating a trivial insulating Ti\,3\textit{d}$^0$ configuration. Nonetheless significant Ti$^{3+}$ spectral weight, chemically shifted to lower binding energies is observed, which we would expect for stoichiometric LTO. The coexistence of both valencies confirms the suspicion that the LaTiO$_3$ is grown overoxidized as LaTiO$_{3+x}$. To determine the $x$ is not easy, as the SrTiO$_3$ substrate also contributes a non-negligible Ti$^{4+}$ signal and in the not unlikely scenario of oxygen vacancies in the substrate also a small Ti$^{3+}$ signal to the spectra.

\clearpage

\section{Sondheimer Oscillations}

\begin{figure*}[h]
	\includegraphics[width=1\textwidth]{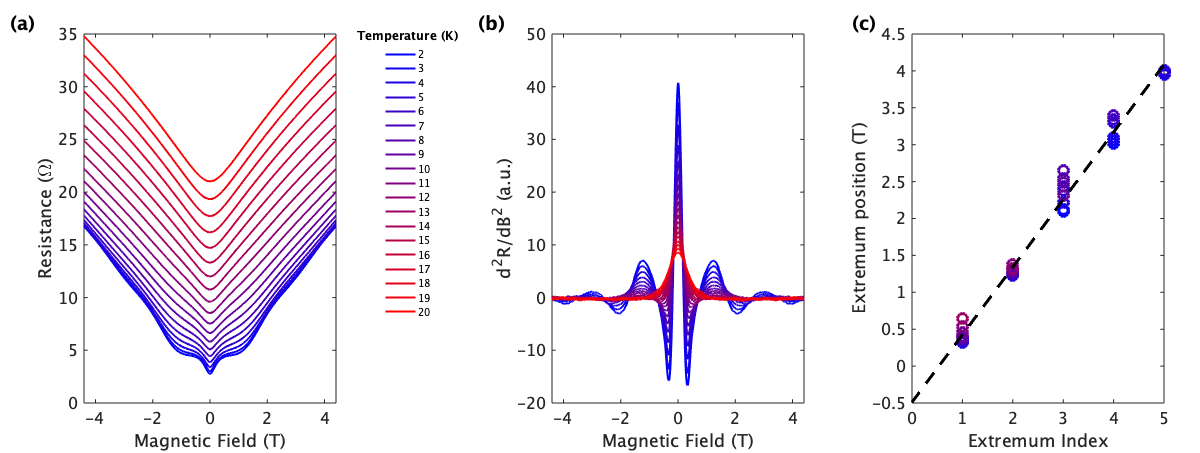}
	\caption{\textbf{Sondheimer oscillations in 3u.c. samples. (a)}, raw longitudinal magnetoresistance at several temperatures. \textbf{(b)}, second derivative of the longitudinal magnetoresistance in \textbf{(a)}, revealing oscillations around zero field at low temperature. \textbf{(c)}, indexed peak position of the oscillations. The linear dependence of the peak position with magnetic field reveals the Sondheimer origin.}
	\label{FigureS_Sondheimer}
\end{figure*}

In 3u.c. thick LTO/STO samples, low-field oscillations were systematically observed (see Fig. \ref{FigureS_Sondheimer}(a)). Due to their small amplitude, those oscillations are more visible in the second derivative of the resistance (see Fig. \ref{FigureS_Sondheimer}(b)). These oscillations are visible at low field, starting from zero field on. Their amplitude is damped with increasing magnetic field, and the oscillations finally disappear above about 4T. The amplitude also decreases in amplitude while temperature is increased, and completely vanishes above about 13K.
\\ \\
The fact that the oscillations exist only at low field and are damped with magnetic field is at odds with that of Shubnikov-de-Haas oscillations, which are often observed in high-mobility thin films but would appear starting from a finite field and increase in amplitude with increasing field. Moreover, the peak position increases linearly with magnetic field, as shown in see Fig. \ref{FigureS_Sondheimer}(c), with a periodicity of $\sim 1$T. This further discounts a Shubnikov-de-Haas origin of these low-field oscillations, which would be periodic in inverse magnetic field.
\\ \\
A possible explanation for those oscillations are the so-called Sondheimer oscillations. They are semi-classical oscillations that are due to a resonance condition between cyclotron radius $r_c$ and thickness of the thin film $t$. When applying a transverse magnetic field to a metallic thin-film, electrons undergo cyclotron orbits in plane, which added to the out-of-plane velocity creates a helical motion. If the mean free path $l_e$ is larger than the film thickness, full helical motion can occur within the film thickness without scattering. The number of cyclotron revolution of the trajectory in the thickness is $n = \tau_z / \tau_c$, with $\tau_z = t/v_z$ the time to propagate in the film thickness from one interface to the other ($v_z$ is the velocity in the out-of-plane direction); and $\tau_c = \frac{2\pi}{\omega_c}$ the time to close a cyclotron orbit in-plane ($\omega_c$ is the cyclotron pulsation). When n is an integer, no net in-plane motion is made while propagating in the thickness. However, when n is non-integer, the helix trajectory is interrupted at the film interface, resulting in a net in-plane motion. As Sondheimer showed, this results in an oscillatory behavior of the resistance with magnetic field, where oscillations occur with period $\Delta B = \frac{\hbar}{et} \frac{\partial A}{\partial k_z} $ \cite{Sondheimer, Harrison1960, vanDelft2021}. Here $\frac{\partial A}{\partial k_z}$ corresponds to the derivative of the cross-section of the Fermi surface $A$ with out-of-plane momentum $k_z$.

The fact that the observed oscillations are periodic in B, along with the fact that the first oscillation occurs at a smaller field than the 1T periodicity (see Fig. \ref{FigureS_Sondheimer}(c)), is compatible with the Sondheimer explanation \cite{Sondheimer}. The extraction of further information about the electronic structure of the film would however require additional information about the band structure and effective mass of the charge carriers, which are unavailable due to the absence of Shubnikov-de-Haas oscillations.

\clearpage

\section{Temperature and field dependence of the linear magnetoresistance}

\begin{figure*}[h]
	\includegraphics[width=1\textwidth]{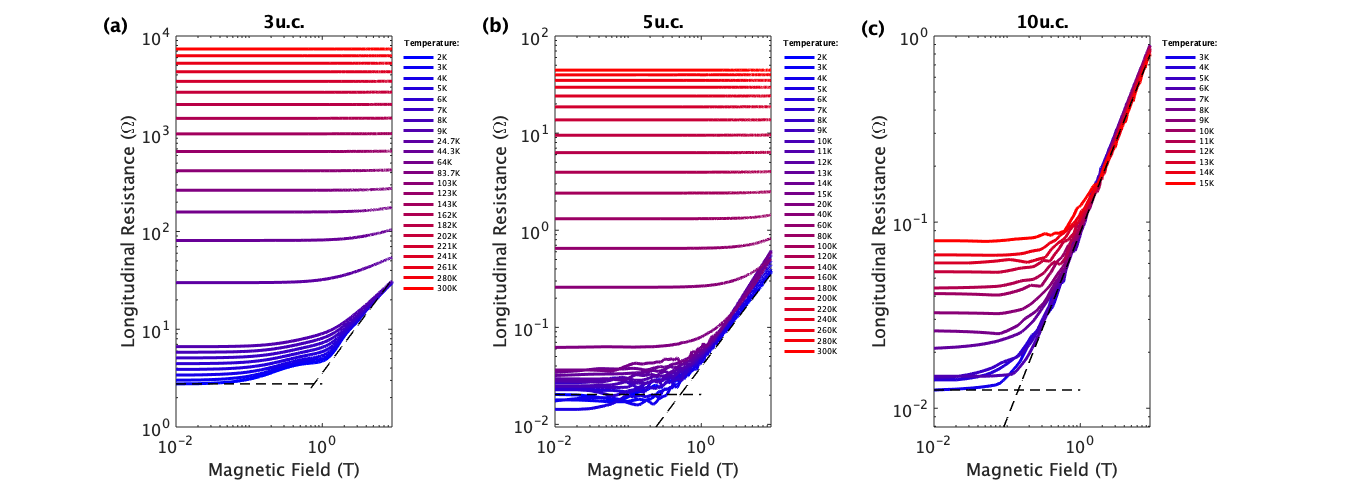}
	\caption{\textbf{Magnetoresistance in log-log scale for samples of all three thicknesses.} For each thickness (\textbf{(a)}: 3u.c., \textbf{(b)}: 5u.c.,\textbf{(c)}: 10u.c.), curves at different temperatures are shown. Dashed lines show linear fits of the lowest temperature data and the intersect with zero-field value.}
	\label{FigureS_RBloglog}
\end{figure*}

In order to determine the onset field of the linear magnetoresistance, we plotted the magnetoresistance in a log-log scale in Fig. \ref{FigureS_RBloglog}. The linear slope in the log-log plot correspond to the linear MR behavior. An estimate of the critical field at which magnetoresistance becomes linear is defined as the intersect between linear fit and zero-field value. We observe that the critical field decreases with increasing thickness. At low temperature, for the 3u.c. sample, this field is close to 1T, while it decrease in the 5u.c. sample to 500mT, and to less than 200mT for the 10u.c. sample (see Fig. \ref{FigureS_RBloglog}).

\begin{figure*}[h]
	\includegraphics[width=1\textwidth]{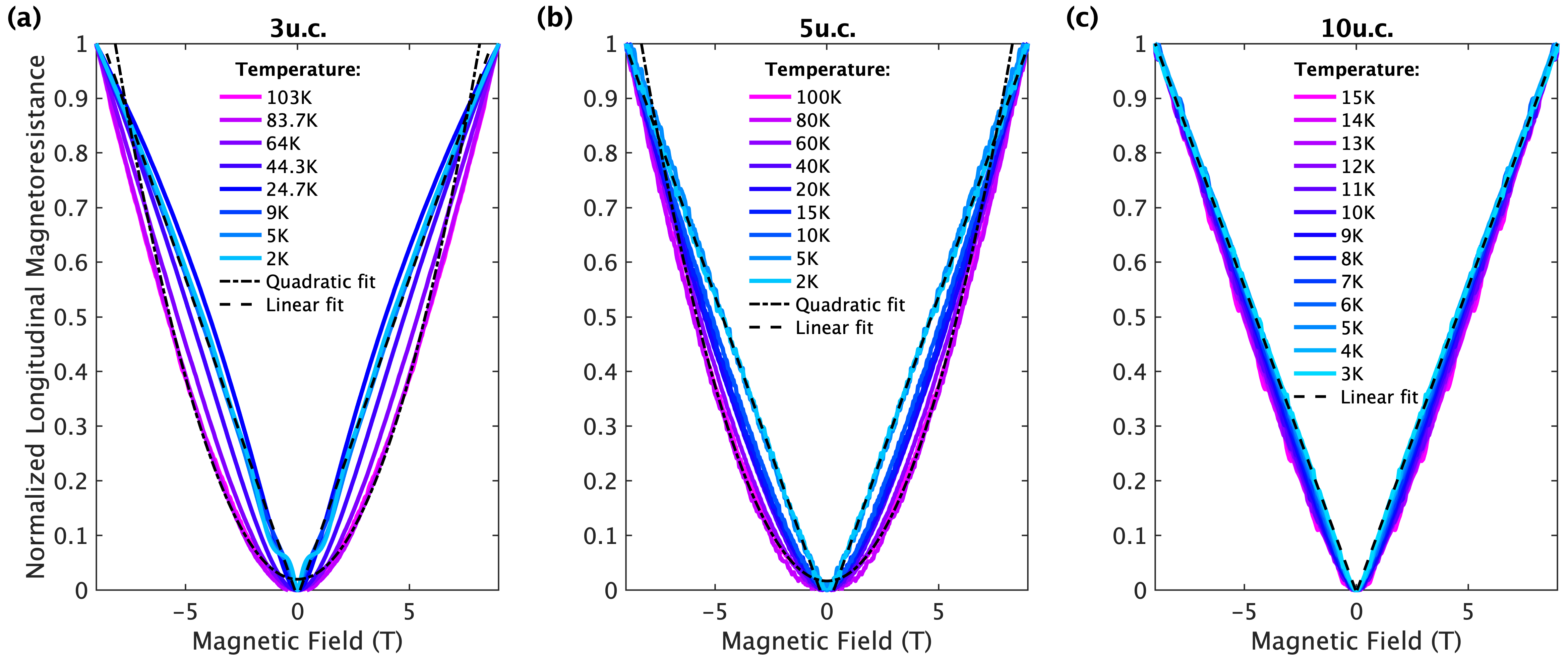}
	\caption{\textbf{Normalised magnetoresistance at various temperatures for samples of all three thicknesses, same as in Fig.\ref{FigureS_RBloglog}.} For each sample (\textbf{(a)}: 3u.c., \textbf{(b)}: 5u.c.,\textbf{(c)}: 10u.c.). Dashed lines are linear fit, and dotted lines are parabolic fits. The transition between linear and parabolic-like behavior occurs around $\sim40$K for the 3u.c. and 5u.c. samples.}
	\label{FigureS_Lineartoparabolic}
\end{figure*}

The linear magnetoresistance studied in the main text is a low-temperature effect. To illustrate this, we plotted in Fig. \ref{FigureS_Lineartoparabolic} the normalized magnetoresistance $$MR_{norm} = \frac{R(B,T)-R(0\textrm{T},T)}{R(9\textrm{T},T)-R(0\textrm{T},T)}.$$ This allows one to observe the transition from a very linear behavior to a parabolic-like behavior. This transition is evidenced in Fig. \ref{FigureS_Lineartoparabolic} by the linear fit at low temperature (dashed line) and the parabolic fits at high temperature (dotted line). For the 3u.c. sample, the transition between both regimes accelerates above ~30-40K, which corresponds to the temperature at which the magnetic stripe pattern observed in cryogenic TEM starts to vanish. For the 10u.c. sample, the magnetoresistance remains linear up to the maximum temperature of 15K.

\clearpage

\section{Lorentz TEM imaging}

\begin{figure*}[h]
	\includegraphics[width=1\textwidth]{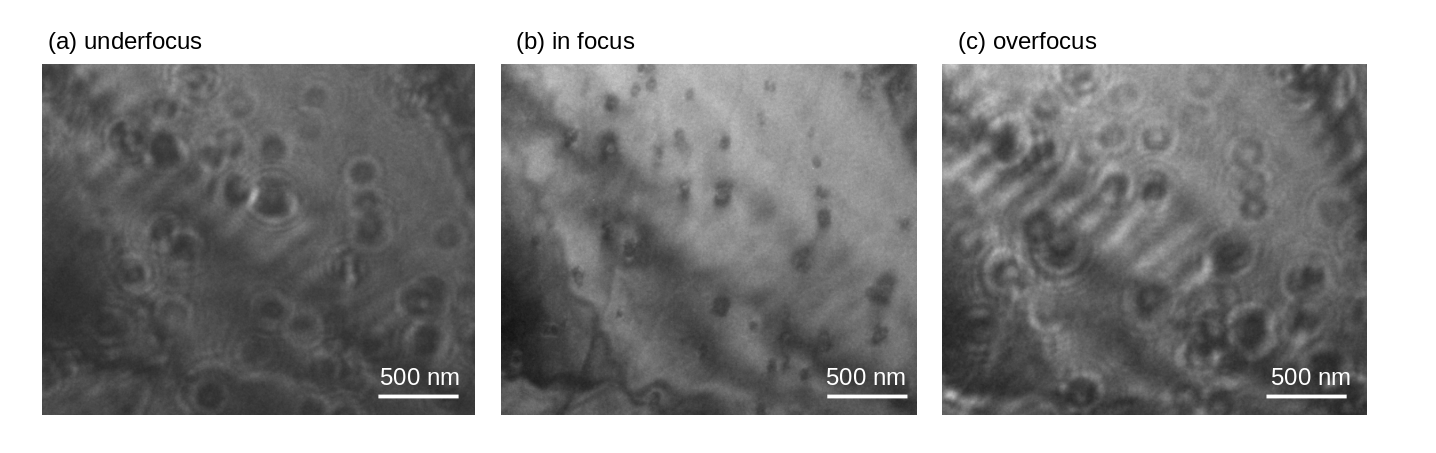}
	\caption{\textbf{(a)} Underfocus, \textbf{(b)} in focus, and \textbf{(c)} overfocus LTEM images of sample region depicted in Fig.3(b,c) in main manuscript. Stripe contrast is inverted going from under- to overfocus and consequently vanished in focus.}
	\label{FigureS_Focal_Series}
\end{figure*}

In this supplement, we supply images of the uniaxial modulation acquired at different foci (Fig. \ref{FigureS_Focal_Series}). They illustrate the fact that the modulation contrast vanished in focus and changes sign going from under- to overfocus, which proves that the observed contrast is a phase contrast (i.e., stems from a phase modulation of the complex electron wave only). Such a modulation typically indicate magnetic textures in the sample. An electric origin (e.g., ferroelectric domains) is ruled out in this case due to the strong magnetic field dependence of the modulation. The small dark dots distributed across the entire field of view may be associated to impurities, or point defects. They are deformed to asterisks due to induced astigmatism caused by the applied external magnetic field.

Note, however, that the small film thickness of the LTO shifts the phase modulation below observability threshold in standard kinematic scattering conditions employed in Lorentz TEM. Typical Lorentz TEM observations of such modulations require magnetic film thickness of several tens of nanometer. As mentioned in the main text, to reveal the magnetic stripes in our thin films we exploit the enhanced sensitivity of dynamic scattering conditions (i.e., scattering beyond first order Born approximation valid close to locally fulfilled Bragg conditions for electron diffraction) towards small phase shift (equivalent to scattering direction) modulations of the electron beam \cite{Lubk2018} imprinted by magnetic fields in the thin \ch{LaTiO3} film (Fig.~3(a) of the main text).
Close to Bragg conditions, where scattering into systematic diffraction directions is strongly enhanced and dynamical scattering sets in (exemplified by the loss of intensity in bending contours of non-diffracted beam), small phase (and corresponding beam direction) modulations can be magnified by more than one order of magnitude depending on the excitation error (i.e., deviation from exact Bragg condition) and other parameters like the thickness of the sample. This is sufficient to amplify the small magnetic phase modulation of the thin LTO film in our case. These are then visualized under strong defocus (LTEM imaging conditions).

\begin{figure*}[h]
	\includegraphics[width=0.5\textwidth]{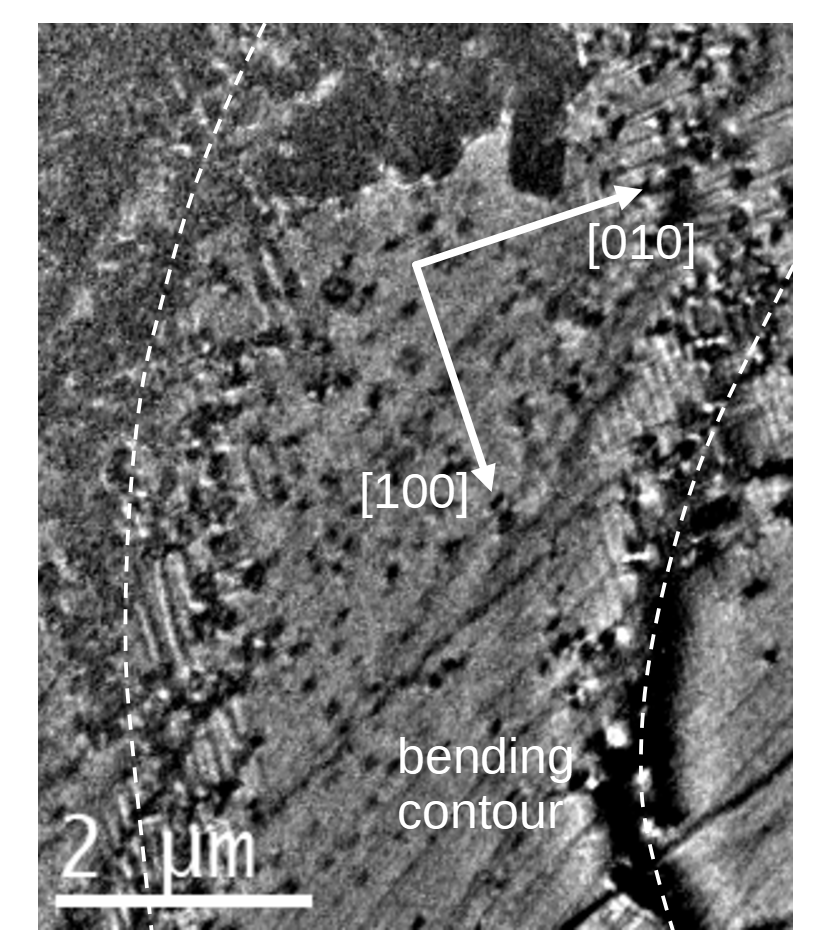}
	\caption{Perpendicular uniaxial stripe modulations aligned with crystal axis orientations of SrTiO$_3$.}
	\label{FigureS_Modulation_Direction}
\end{figure*}

Lastly, we provide in Fig. \ref{FigureS_Modulation_Direction} another example of uniaxial stripe modulations at bending contours, which exhibit two perpendicular modulation directions aligned with the crystallographic orientations of the underlying SrTiO$_3$ substrate in one field of view.

\clearpage

\section{Theoretical modeling}
\subsection{Magnetism in the \ch{LaTiO3} heterostructure}
\begin{figure*}[ht]
\centering
	\includegraphics[width=0.4\textwidth]{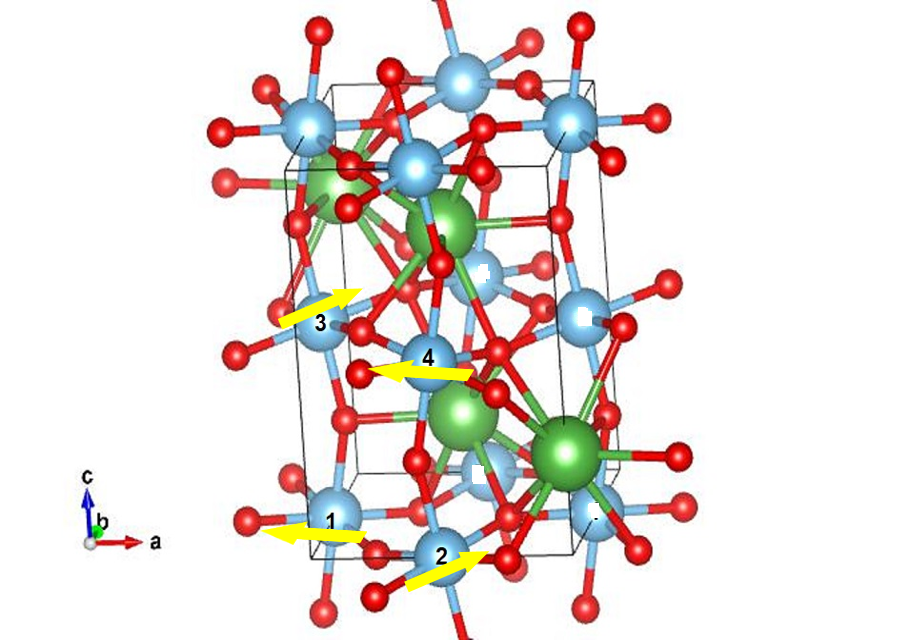}
	\caption{\textbf{Magnetic moments (yellow arrows) on the four primitive Ti$^{3+}$ ions in the LaTiO$_3$ structure. They are labelled 1 - 2 -3 -4.Ti atoms are shown in blue, O atoms in red and La atoms in green.}}
	\label{Figure5}
\end{figure*}

Experimental \cite{Meijer1999} and theoretical \cite{Schmitz2005} investigations established that bulk LaTiO$_{3}$ orders antiferromagnetically (AF) at low temperature  with canted Ti$^{3+}$ moments, resulting in a small ferromagnetic component. The structure is shown in Fig. \ref{Figure5}, where $(a,b,c)$ refers to the orthorhombic basis vectors. $\theta \sim 1^{\circ}$ is the  angle of the magnetic moments with respect to the $ (a,b)$ plane and $\phi \sim 1.4^{\circ}$ denotes the absolute value of the angle between the projection of the moments onto the $ (a,b)$ plane and the $a$ axis.  The canted AF ordering is caused by Dzyaloshinskii-Moriya (DM) and exchange asymmetry terms allowed by symmetry. Doping the system with holes leads to a percolative-style decrease in the magnitude of the zero temperature average moment as observed in experiments on LaTiO$_{3+\delta}$ samples with  $0 \le \delta \le 0.08$ \cite{Meijer1999}. At low temperature, a metal-to-insulator transition occurs at a critical doping $\delta \sim 0.05$ such that the material is insulating for $\delta < 0.05$, while it hosts coexisting metallic and magnetic states for $\delta > 0.05$. The off-stoichiometry oxygen holes dope the Ti ions turning a fraction of them into Ti$^{4+}$. This causes site dilution of the magnetic Ti$^{3+}$ moments and also produces metallic puddles. When $\delta \sim 0.05$ puddles merge to form a percolative metallic phase \cite{Kurdestany2017}. When $\delta > 0.08$  a homogeneous metallic and paramagnetic state is obtained.
\\ \\

\indent The LTO/STO heterostructure is polar and in order to mitigate the electrostatic energy build-up \cite{Nakagawa2006} an electron liquid is produced in the material. If we consider that the negative charges originate from the Ti$^{3+}$ ions on the LTO side, we may transpose the previous considerations to the interface case. We note that, in contrast to the LaAlO$_3$/SrTiO$_3$ case, metallicity is expected to concern both sides of the interface. If the uniaxial modulations observed in TEM are caused by a spiral magnetic structure, as we argue in the main text, we conclude that in our experiments, the electron fraction $n_f$ donated by each Ti$^{3+}$ site of LTO in the heterostructure is such that  $0.1<n_f<0.16$. In the percolative metallic phase, the volume fractions $v_M$ of the metal and $v_I$ of the insulator satisfy $v_M\; + v_I = 1$ and $0.16\;v_M =n_f$ (Maxwell construction in a two-phase state). At the percolation threshold $v_M \sim 0.62$, which is close to the critical occupation for  the square lattice site percolation problem. The very low value of the resistivity of the 10 u.c. samples suggests a proximity to the homogeneous phase and a sizable carrier concentration, so we estimate $n_f= 0.14$, i.e $v_M = 88$\%. 

In the effective Hamiltonian $\cal{H}_{\textrm{eff }}$ (equation (2) of the main text), the anisotropy tensor $A$ has non-zero elements in the plane of the interface. Due to the tilt and rotation of the TiO$_6$ octahedra in LTO, hopping amplitudes are given by a non-diagonal matrix\cite{Schmitz2005} such that the interfacial DM mostly affects $\phi$.
\\ \\
\indent To see the stability of the spiral state, we checked that the demagnetizing energy due to the small ferromagnetic component at the boundaries \cite{Gabay1985} does not alter the spiral magnetic ground state. The stability of the spiral, when a magnetic field is applied perpendicularly to the interface, depends on the strength of the in-plane DM of LTO and of the exchange asymmetry. Using values listed in Ref (\onlinecite{Schmitz2005}), with a typical anisotropy parameter $A\simeq 2$meV, we estimate that moments align along $z$ when the magnetic field $B$ exceeds $\sim 20$T, which is above the maximum field available in our magnetotransport measurements.

\newpage

\subsection{Mobility in the spin spiral domains}
 As can be seen in Fig. 3 of the main text, the linear scale of the magnetic modulation region is on the order of a micrometer. According to the scenario that we advocated to explain its origin, this suggests that the spin diffusion length, $\lambda_s$, has a similar magnitude. As electrons move through the stripe pattern, they experience spin scattering due to the change in the direction of alignment of the local moments, perpendicularly to the stripe. Following \cite{MacDonald2006}, we determine the transmission $T$ and reflection $R$ coefficients {\it a la } Landauer-B\" uttiker, for electrons traveling across a structure consisting of two antiferromagnetically ordered regions separated by a paramagnetic spacer layer. The direction of the alignment  in the second region is at an angle $\theta$ with respect to that in the first region. The system is equivalent to a Fabry-P\' erot interferometer, so that we model one period of the spin spiral pattern as  $d/\mathbf{a_0}$ interferometers in series ($d= 160$ nm, $\mathbf{a_0}= 4\; \AA$). A priori one should include two main types of processes. In the direct process, a portion $T$ of the wave hits interferometer $ n$ and is subsequently transmitted to $n+1$. In the indirect process,  the portion $R$ of the wave that is reflected from interferometer $n$, travels in the backward direction towards interferometer $m$, where part of it is reflected and then moves in the forward direction towards interferometer $n$ where it adds its contribution to the transmitted portion $T$ of the initial wave; the amplitude of this indirect process is at most $T^{n-m}R^2$, which decreases exponentially with $(n-m)$; in addition it contains a phase factor, which stems from the $2(n-m)a_0$ distance involved in the indirect process, producing a phase difference between the two amplitudes. Overall, we neglect these secondary processes, which introduce small corrections to $T$ and $R$.
For $\theta = \frac{2\pi}{d}\mathbf{a_0} \sim 0.016$, $T\sim 0.999$ \cite{MacDonald2006} and over a distance $\lambda_s \sim 1\;\mu$m, there are  $\frac{\lambda_s}{\mathbf{a_0}}$  interferometers. The ratio of the mobility in the presence of the spin spiral over the mobility in its absence is $ T^{\lambda_s/\mathbf{a_0}}/(1-T^{\lambda_s/\mathbf{a_0}} \sim 0.05$.
Consequently the mobility in the stripe region is at least one order of magnitude smaller than in the rest of the sample.

\end{document}